\documentclass[aps,prb,twocolumn,superscriptaddress,nofootinbib]{revtex4-2}
\usepackage{amssymb,graphicx,color}
\usepackage[intlimits]{amsmath}
\usepackage[english]{babel}
\usepackage[colorlinks]{hyperref}
\hypersetup{
colorlinks=blue,
linkcolor=blue,
filecolor=blue,
urlcolor=blue,
citecolor=blue
}
\usepackage[normalem]{ulem}
\usepackage{comment}
\usepackage{ragged2e}
\usepackage{enumerate}
\usepackage{enumitem}

\graphicspath{{figures/}} 
\usepackage{multirow}
\usepackage{color, colortbl}
\usepackage[first=0,last=9]{lcg}
\newcolumntype{P}[1
]{>{\centering\arraybackslash}p{#1}}

\definecolor{col2}{RGB}{220,223,244}
\definecolor{utred}{RGB}{165, 30, 55}
\definecolor{col3}{RGB}{205, 220, 244}
\definecolor{cola}{RGB}{221, 220, 248}
\definecolor{col4}{RGB}{250,209,220}
\definecolor{col0}{RGB}{244, 225, 232}
\definecolor{col1}{RGB}{240,229,240}

\begin{document}

\title{\hspace{-0pt} Periodically and aperiodically Thue-Morse driven
  long-range systems: from dynamical localization to slow dynamics}
\author{Vatsana Tiwari}
\affiliation{Department of Physics, Indian Institute of Science Education and Research, Bhopal, India}
\author{Devendra Singh Bhakuni}

\affiliation{The Abdus Salam International Centre for Theoretical Physics (ICTP), Strada Costiera 11, 34151 Trieste, Italy}
\author{Auditya Sharma}
\email{auditya@iiserb.ac.in}
\affiliation{Department of Physics, Indian Institute of Science Education and Research, Bhopal, India}

\begin{abstract}

We investigate the electric-field driven power-law random banded
matrix(PLRBM) model where a variation in the power-law exponent
$\alpha$ yields a delocalization-to-localization phase transition. We
examine the periodically driven PLRBM model with the help of the
Floquet operator. The level spacing ratio and the generalized
participation ratio of the Floquet Hamiltonian reveal a drive-induced
weak multifractal (fractal) phase accompanied by diffusive (subdiffusive) transport on the delocalized
side of the undriven PLRBM model. On the localized side, the
time-periodic model remains localized - the average level-spacing ratio
corresponds to Poisson statistics and logarithmic transport is
observed in the dynamics. Extending our analysis to the aperiodic
Thue-Morse (TM) driven system, we find that the aperiodically driven
clean long-range hopping model (clean counterpart of the PLRBM model)
exhibits the phenomenon of \textit{exact dynamical localization} (EDL)
on tuning the drive-parameters at special points. The disordered
time-aperiodic system shows diffusive transport followed by relaxation
to the infinite-temperature state on the delocalized side, and a
prethermal plateau with subdiffusion on the localized side.
Additionally, we compare this with a quasi-periodically driven AAH
model that also undergoes a localization-delocalization
transition. Unlike the disordered long-range model, it features a
prolonged prethermal plateau followed by subdiffusion to the infinite
temperature state, even on the delocalized side.

\end{abstract}

\maketitle

\section{Introduction}

Recent experimental evidence of non-equilibrium
phases~\cite{schreiber2015observation,Choi2017observational,guo2020stark,
  Guo2021,morong2021observation,Mi2022,dotti2024measuring} has
underscored the significance of Floquet engineering and disorder as
mutually competitive and essential tools for controlling and
manipulating quantum systems via creation of optical lattices and band
structures~\cite{controlling2011chen,Rechtsman2013photonic,bukov2015universal,holthaus2015floquet,
  Schindler2024counterdiabatic}, artifical gauge
fields~\cite{Lin2009,Periodically2014goldman,Roushan2017chiral,Rosen2024synthetic},
topological charges~\cite{martin2017topological,boosting2022long,Citro2023thouless} and photon
pumps~\cite{photon2021christina,martin2017topological,topological2019crowley}. Both
the tools facilitate non-trivial quantum transport and the emergence
of novel phases that are absent in equilibrium
systems~\cite{anderson1958absence,aubry1980analyticity,
  harper1955single,holthaus1995ac}. The phenomenon of Anderson
localization observed in randomly disordered and quasiperiodic
disordered systems has generated a voluminous literature~\cite{anderson1958absence,disordered1985Lee}.
The Aubry-Andr{\'e}-Harper (AAH) model is the paradigmatic example of
quasiperiodic systems where a non-zero disorder strength is required
to enforce single-particle
localization~\cite{anderson1958absence,harper1955single,aubry1980analyticity}
even in one-dimension. The tilted potential in its static and
time-dependent forms represents an alternate disorder-free mechanism
for localization via the phenomena of Wannier-Stark
localization~\cite{zener1934theory,Wannier1960wave,krieger1986time,Guo2021},
and dynamical
localization~\cite{dunlap1986dynamic,dunlap1988dynamic,Dignam2002conditions,eckardt2009exploring,
bhakuni2018characteristic,tiwari2022noise,tiwari2024dynamical} respectively.
\begin{figure}
\centering
\hspace{2ex}\includegraphics[width=0.45\textwidth]{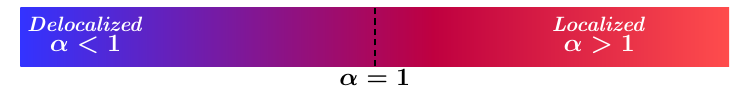}
\hspace{-17ex}\includegraphics[width=0.45\textwidth]{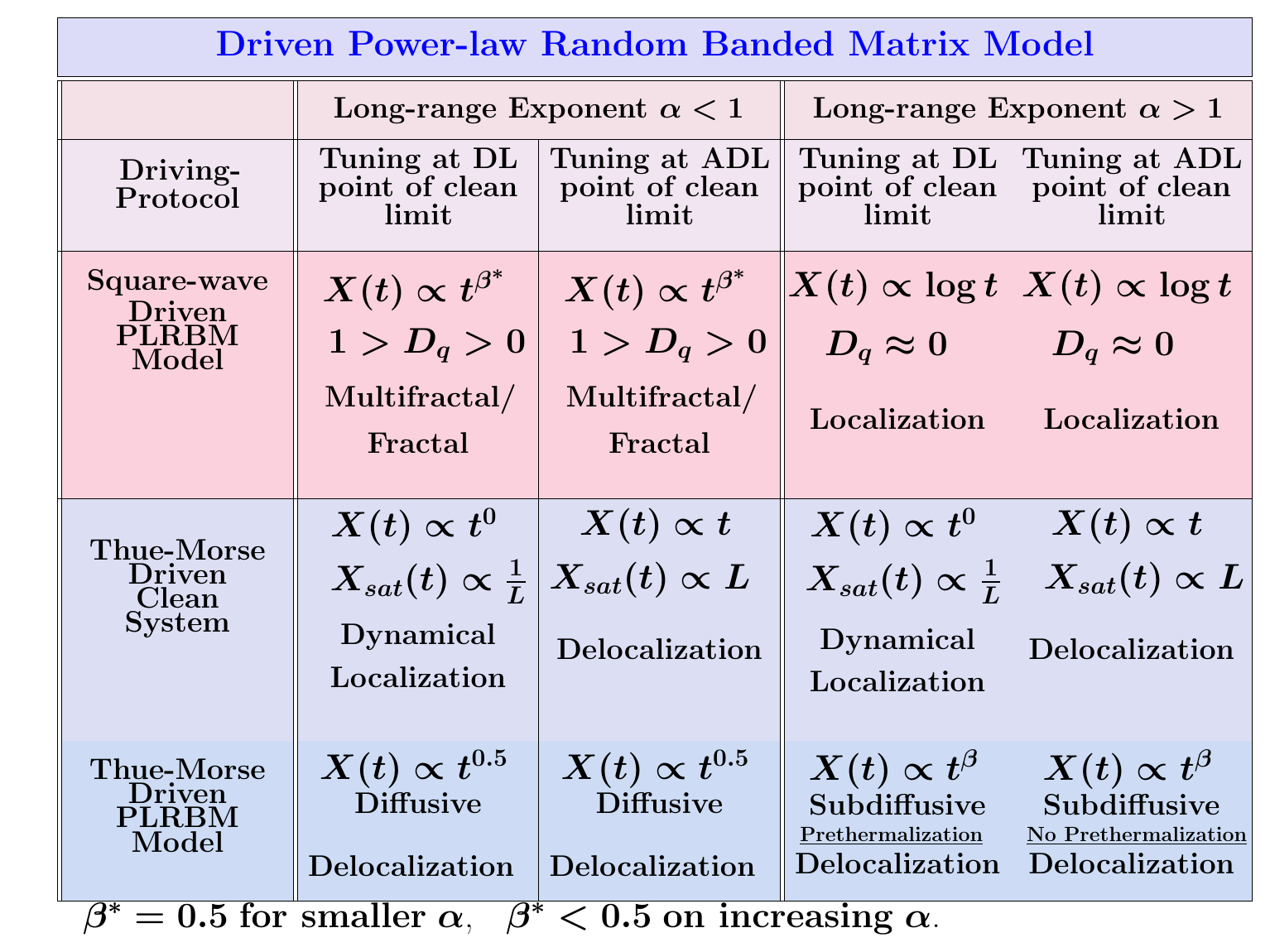}
\caption{Schematic representation of our main findings and comparison
  with the undriven PLRBM model shown in the rectangluar color bar. In
  the table, we present the results of periodically and aperiodically
  driven PLRBM model based on our analysis of RMSD ($X(t)$),
  entanglement entropy ($S(t)$), and fractal dimension ($D_{q}$). In the first row, the power-law exponent $\beta^{*}=0.5$ corresponds to diffusive transport for smaller $\alpha$, and $\beta^{*}<0.5$ corresponds subdiffusive transport as $\alpha$ increases(Fig.~\ref{fig:fig2b}). Multifractal and fractal behvaior is confirmed from the study of $D_{q}$ as a function of $q$ for different values of long-range exponents $\alpha $ (Fig.~\ref{fig:fig1b}). In
  the bottom-most row, $\beta<0.5$ corresponds subdiffusive
  transport.}
\label{fig_table}
\end{figure}
Excitingly, Floquet engineering of disordered systems allows for the
realization of exotic features such as drive-induced Anderson
localization~\cite{holthaus1995ac,holthaus1995random}, drive-disorder
dependent localization-delocalization phase
transition~\cite{roy2015fate,abanin2016theory,bairey2017driving,dotti2024measuring},
and Stark many-body
localization~\cite{schulz2019stark,morong2021observation,
  taylor2020experimental,zisling2022transport,liu2022discrete,zhang2023floquet,stark2024duffin}. Expanding
the landscape of driving protocols beyond the Floquet
setting~\cite{nandy2018steady,dumitrescu2018logarithmic,zhao2019floquet,else2020long,long2022many},
structured aperiodic driving allows for not only dynamical
localization but also remnants of non-equilibrium features in the form
of slow logarithmic relaxation~\cite{tiwari2024dynamical}. In contrast
to periodic driving, aperiodic driving also encompasses a wide-range
of phenomena such as coherence
restoration~\cite{mukherjee2020restoring},
localization~\cite{long2022many,zhao2022localization},
prethermalization~\cite{mori2021rigorous,zhao2022suppression,dutta2024prethermalization},
the occurrence of time
quasicrystals~\cite{dumitrescu2018logarithmic,else2020long}, and
complete Hilbert-space ergodicity (CHSE) owing to the irreducibility
of time-quasiperiodic
dynamics~\cite{complete2023Pilatowsky,pilatowsky2024hilbert}.

A large number of experimental realizations such as certain quantum
simulators e.g., ion-traps~\cite{Smith2016,morong2021observation},
Rydberg atoms~\cite{Jae2016exploring}, dipole-dipole
interactions~\cite{Choi2017observational,Kucsko2018critical,Davis2023},
and nitrogen-vacancy centers in
diamond~\cite{Choi2017observational,Kucsko2018critical,Zhou2020quantum,Zu2021emergent,Davis2023}
are characterized by long-range coupling, typically modeled as a
power-law decaying function ($1/r^{\alpha}$).  Remarkably, the
interplay of driving and long-range coupling can yield stable
localization in the non-interacting limit, while in the interacting
limit, broken algebraic MBL~\cite{Burau2021Fate},
Floquet/quasi-Floquet
prethermalization~\cite{machado2020long,peng2021floquet,bhakuni2021suppression,He2023quasi},
and exponentially slow thermalization~\cite{machado2019exponentially}
have been reported.  In this work, we explore the interplay of
power-law hopping $(1/r^{\alpha})$ and time-periodic and aperiodic
electric field in the driven power-law random banded matrix (PLRBM)
model~\cite{transition1996Mirlin,anomalous2001cuevas,fluctuations2000evers}. The
colour bar shown in Fig.~\eqref{fig_table} represents the static PLRBM
model which is known to exhibit three phases with long-range exponent
$\alpha$: delocalized phase for $\alpha<1$, multifractality at
$\alpha=1$, and localized phase for $\alpha>1$. In the table
(Fig.~\eqref{fig_table}), we summarise our main findings based on our
analysis of RMSD $X(t)$, entanglement entropy $S(t)$, and fractal
dimenstion $D_{q}$. Our study reveals that an electric field drive
induces a suppression in the range of hopping as a result of the
renormalization of the hopping strength. The electric field drive thus
has a tendency to shift the system from longer-range hopping towards
an effective shorter range hopping, which results in a weakly
delocalized fractal phase on the delocalized side of the static model.

We begin our study by considering a time-periodic long-range hopping
model and evaluate the corresponding Floquet Hamiltonian, recovering
the well-known phenomenon of \textit{exact dynamical localization}
(EDL)~\cite{dunlap1986dynamic,Dignam2002conditions,dynamic2002domachuk,dynamic2004wan,
  eckardt2009exploring,quasi2009joushaghani,
  generalized2012joushaghani}. Next, we extend this analysis to the
study of the disordered hopping model, where dynamical localization
(DL) is destroyed by the disorder. To characterize the drive-induced phases as the long-range exponent $\alpha$ varies, we employ static measures such as the level spacing ratio and the generalized inverse participation ratio. These metrics uncover a weak multifractal phase, which transitions to a fractal phase on increasing $\alpha$ on the delocalized side of the undriven PLRBM model, and a localized phase on the other side of the transition.  Furthermore, the dynamics of the periodically driven system reveals that the transport changes from diffusive to subdiffusive as $\alpha$ is increased on the delocalized side of the undriven model, and exhibits logarithmically slow transport on the localized side of the transition.

Under aperiodic driving, the clean system with long-range hopping
shows the emergence of \textit{exact dynamical localization}, while
showing ballistic transport at away from dynamical localization (ADL)
points. In the disordered case, the system exhibits diffusive
transport to the infinite temperature state on the delocalized side of
the PLRBM model, and a prethermal plateau followed by diffusive
transport to the infinite-temperature state on the localized
side. Finally, we include a short discussion on the quasiperiodically
driven Aubry-Andr{\'e}-Harper model~\cite{
  aubry1980analyticity,harper1955single}, which also features a
delocalization-localization transition. In contrast to the delocalized
side of the disordered long-range model, we observe that the aperiodic
drive suppresses the transport and gives rise to distinct dynamical
regimes-- a prethermal plateau for a long time followed by
subdiffusive growth to the infinite temperature state at late times.

The organization of the paper is as follows: In section~\ref{Model
  Hamiltonian}, we introduce the model Hamiltonian, the driving protocols
and the observables used to study the dynamics. We then discuss the results
of the periodically driven PLRBM model in section~\ref{Periodically Driven
  PLRBM Model}, followed by the results of the aperiodic Thue-Morse
driven system in section~\ref{Aperiodic Thue-Morse Driving}. We finally
discuss and conclude our findings in section~\ref{Summary}.

\section{Model Hamiltonian, driving protocols and observables} \label{Model Hamiltonian}  

We consider a one-dimensional disordered long-range hopping fermionic
chain subjected to a time-dependent electric field. The model
Hamiltonian can be written as
\begin{align}
\label{eqn:eq1a}
H(t)=-\sum_{i,j=1}^{L-2}\frac{J_{ij}}{|i-j|^{\alpha}}\left(c_{i}^{\dagger}c_{j}+ h.c.\right)  + \mathcal{F}\left(t\right)\displaystyle \sum_{j=0}^{L-1}jn_{j}.
\end{align}
Here, $J_{ij}= (J + u_{ij}) $ is the hopping strength with $ u_{ij}$
being random numbers drawn from a uniform distribution in the interval
$ [-1,1] $, $\mathcal{F}(t)$ is the time-dependent driving strength,
and $\alpha$ is the long-range parameter. In the absence of the
time-dependent field ($\mathcal{F}\left(t\right)=0$), the Hamiltonian
is the well-known power-law random banded matrix (PLRBM) model which
features a delocalization to localization transition at
$\alpha=1$~\cite{transition1996Mirlin,anomalous2001cuevas,fluctuations2000evers,roy2018entanglement}. For
$\alpha<1$, all the eigenstates are delocalized whereas for
$\alpha>1$, all the eigenstates are localized.

In the presence of the time-dependent drive, and specifically for a
periodic drive: $\mathcal{F}(t+T)=\mathcal{F}(t)$, the clean limit
($u_{ij}=0$) exhibits the phenomenon of exact dynamical localization for
some specific choice of the driving amplitude and the frequency where
the dynamics features revivals with the driving frequency and
consequent absence of transport~\cite{dunlap1986dynamic,Dignam2002conditions,dynamic2002domachuk}. In
this work, we focus on the effect of such a time-dependent drive on
the disordered long-range model and extend it for the case of an
aperiodic drive. For the periodic drive, the time-dependent electric
field $\mathcal{F}(t)$ oscillates periodically between $\pm F$ with
the time-period $T$,
\begin{eqnarray}
\mathcal{F}\left(t\right)=  \left\{\begin{array}{cc} +F,  & 0\leq t \leq T/2  \\
- F,  & T/2<t \leq T \end{array}\right.. \label{eqn:eq3} 
\end{eqnarray}
For the aperiodic drive, we consider the Thue-Morse protocol~\cite{zhao2021random,tiwari2024dynamical}. The Thue-Morse sequence (TMS) can be generated using the recurrence relation
\begin{eqnarray}
	\label{TMS_U}
	U_{n+1}=\tilde{U_{n}}U_{n}, \quad \tilde{U}_{n+1}&=&U_{n}\tilde{U_{n}},
\end{eqnarray}
where we start with the unitary operators $U_{1}= U_{-}U_{+}$, $\tilde{U_{1}}= U_{+}U_{-}$
with $U_{-}=e^{-iTH_{B}}$, and $U_{+}=e^{-iTH_{A}}$. The time evolution of any initial wave function is given by : $|\psi(2^{n}T)\rangle = U_n|\psi\left(0\right)\rangle$. 
\begin{figure}[t]
\centering
\hspace{-2ex}\includegraphics[width=0.47\textwidth]{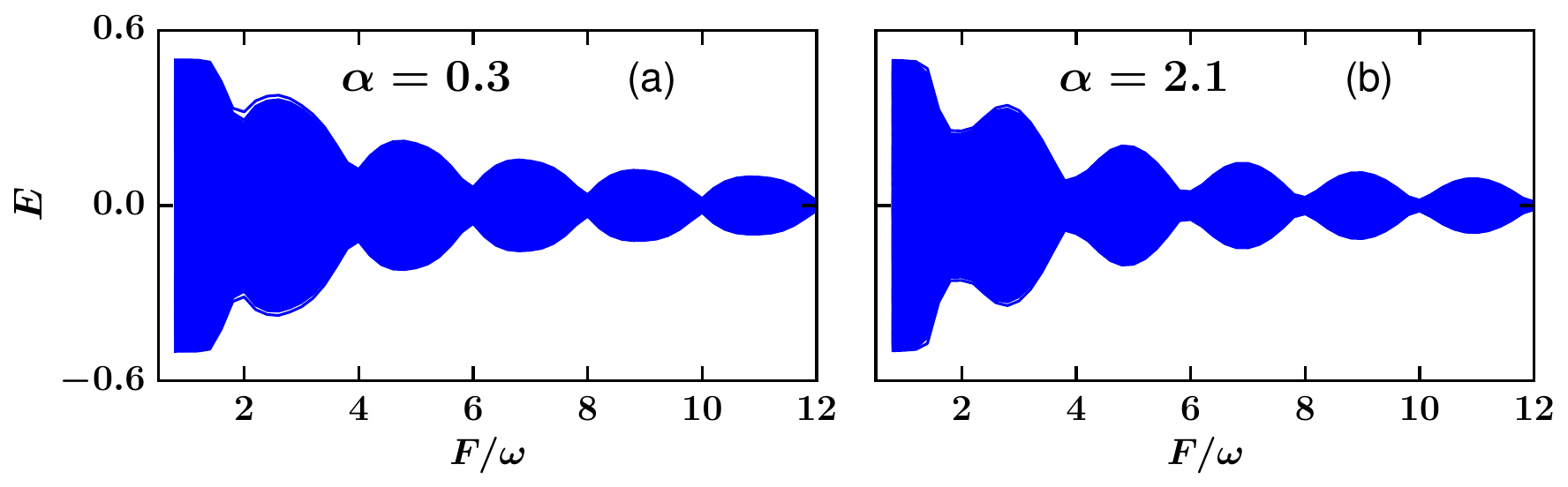}
\caption{Quasienergy spectrum of periodically driven PLRBM (Floquet PLRBM) for long-range exponents $\alpha=0.3, 2.1$. The other parameters are system-size $L=500,$ and driving-frequency $\omega=1$.}
\label{fig:QE}
\end{figure}

\begin{figure*}[t]
\centering
	\hspace{-2ex}\includegraphics[width=1\textwidth]{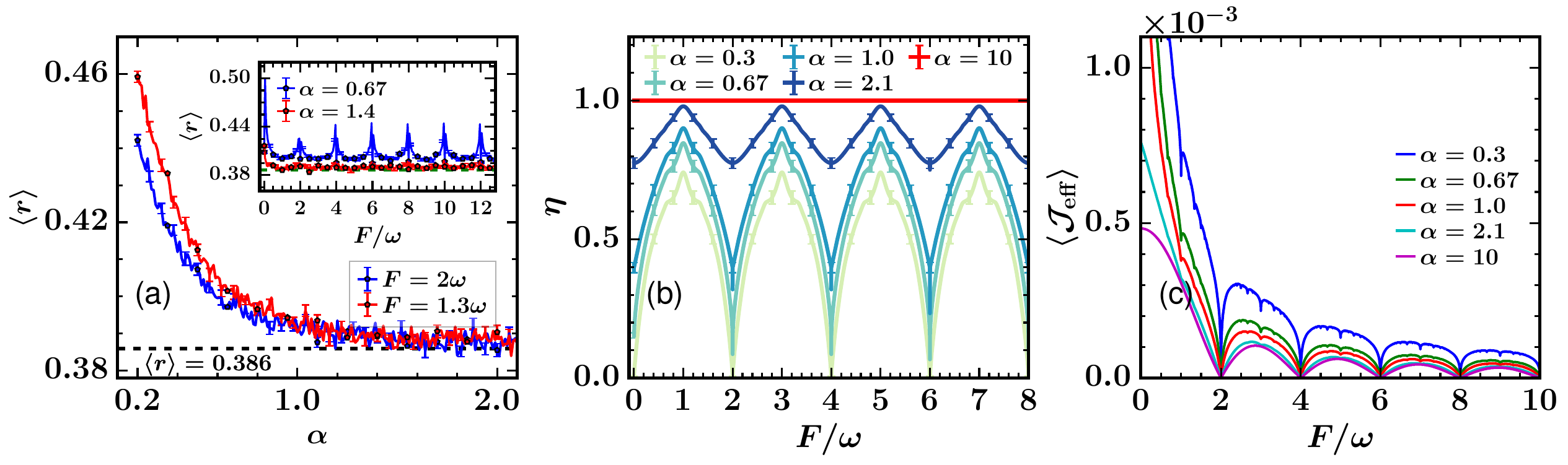}
	\caption{(a) Average gap ratio $\langle r\rangle $ with long-range exponent $\alpha$ for the driving-parameters tuned at drive-amplitude $F=1.3\omega, F=2\omega$. Inset(a). Average gap ratio $\langle r\rangle $ with drive-parameters $F/\omega$ for long-range exponent $\alpha=0.67,1.4$. The other system parameters are driving-frequency $\omega=2$, and system-size $L=1024$. The black dashed line corresponds to the average gap ratio for Poisson statistics, $\langle r\rangle=0.386$. (b) $\eta $ obtained from $\mathcal{J}^{\text{eff}}_{p}$ with drive-parameters $F/\omega$ for different long-range exponents. (c) Average hopping $\langle \mathcal{J}_{\text{eff}}\rangle$ with drive-parameters $F/\omega$ for different long-range exponents for system size $L=1024$. We have performed average over 100 disorder realizations for all the shown data. }
	\label{fig:fig1a}
\end{figure*}

To study the dynamics and the nature of underlying transport, we focus
on the study of root-mean-squared displacement of a wave packet
$|\psi(0)\rangle $ initially localized at the central site ($i_{0}=L/2$) of the
one-dimensional fermionic chain,
\begin{equation}
\label{eq:msd}
X(t)= \biggl[\biggl(\sum_{i=0}^{L}(i-i_{0})^{2}|\psi_{i}(t)|^{2}\biggr)\biggr]^{1/2}.
\end{equation} 
The dynamics of the root
mean-squared displacement (RMSD) $X(t)$ typically grows as a power
law: $X(t)\sim t^\beta$, and the different transport regimes are
distinguished by the exponent $\beta$. For ballistic transport
$\beta=1$, whereas for diffusion $\beta=1/2$. $\beta>1/2$ and
$\beta<1/2$ correspond to superdiffusion and subdiffusion
respectively. For localized systems, $\beta=0$ corresponds to no
transport.

In addition to RMSD, we study the half-chain entanglement entropy from
an initial Neel state: $|\psi(0)\rangle =\prod_{i}^{L/2}
c^{\dagger}_{2i}|0\rangle $. The entanglement-entropy of a subsystem
$A$ is given by
\begin{equation}
	\label{Sent}
	S(t) = -\text{Tr}[ \rho_A(t) \text{ln} \rho_A(t)].
\end{equation}
Here, $\rho_A(t)=\text{Tr}_{B} \rho(t),$ where $\rho(t)=|\psi(t)\rangle \langle \psi(t)|$ is the density-matrix of the system, and $\rho_{A}(t)$ is the reduced density matrix of the subsystem $A$ obtained after tracing out the other part $B$ of the subsystem. For non-interacting systems, the entanglement entropy can be calculated from the eigenvalues $\lambda_\alpha$ of the correlation matrix as~\cite{peschel2003calculation,entanglement2006Gabriele,roy2018entanglement} 
\begin{equation}
	S(t) = - \sum_{\alpha}[ \lambda_\alpha(t) \text{ln} \lambda_\alpha(t) + (1-\lambda_\alpha(t)) \text{ln} (1-\lambda_\alpha(t))].
\end{equation}
Many studies have shown that for single particle localization, $S(t)$
saturates to a constant value, and for clean systems featuring
ballistic transport, $S(t)$ shows linear growth~\cite{Kim2013ballistic}. However, a sublinear
growth of entanglement entropy is seen in the case of anomalous
transport~\cite{luitz2017ergodic,tiwari2024dynamical}.

\section{Periodically Driven PLRBM Model}
\label{Periodically Driven PLRBM Model}
We first consider the case of a periodically driven system, where the
dynamics is governed by an effective Hamiltonian, which can be
obtained from the one-cycle unitary operator known as the Floquet
operator:
\begin{equation}
	U(T)=e^{-iH_{A}T/2} e^{-iH_{B}T/2}\equiv \exp(-iTH_{\text{F}}),
\end{equation}
where $H_{A/B}$ are the Hamiltonians for the two-cycles with field
strength $\pm F$, and $H_{\text{F}}$ is the effective
Hamiltonian. First, we discuss the clean limit of Eq.~\eqref{eqn:eq1a}
($J=1, \ u_{ij}= 0$) which is well-known to exhibit the phenomenon of
exact dynamical localization. The phenomenon of dynamical localization
can be understood with the help of a Magnus expansion of the clean
counterpart of the time-periodic Hamiltonian
(Eq.~\eqref{eqn:eq1a}). In this case, the model Hamiltonian can be
defined as
\begin{eqnarray}
  H_{A/B}&=& -\sum_{p>0}\frac{J}{p^{\alpha}}\left(\hat{K}_{p}+\hat{K^{\dagger}_{p}}\right)\pm F\sum_{p=0}^{L-1}jn_{p},
\end{eqnarray}
where we define the unitary operators as~\cite{hartmann2004dynamics}
\begin{eqnarray}
  \label{eq:k}
  \hat{K}_{p}=\sum_{n}c_{n}^{\dagger} c_{n+p},\quad   \hat{N}=\sum_{n}nc_{n}^{\dagger}c_{n}.
\end{eqnarray}
Following the Baker-Campbell-Hausdorff formula~\cite{Hall2015}, we can
obtain an effective Hamiltonian as
\begin{eqnarray}	
\label{Eq.Hf1}
H_{\text{eff}}&=& -\sum_{p>0}\mathcal{J}_{p}^{\text{eff}}\left(\hat{K}_{p}e^{-iFT/4p}+\hat{K}^{\dagger}_{p}e^{iFT/4p}\right),
\end{eqnarray}
where, $\mathcal{J}_{p}^{\text{eff}}=\frac{J\sin
  \left(pFT/4\right)}{p^{\alpha}\left(pFT/4\right)},$~\cite{Dignam2002conditions,eckardt2009exploring,
  bhakuni2020stability}.  To obtain the condition for \textit{exact
  dynamical localization} (EDL), we require that the renormalized
hopping-strengths corresponding to all $p$'s vanish simultaneously,
and it happens at $F=2m\omega$, where $m\in \mathbb{Z}$. It is worth
noting that the phenomenon of EDL emerges under the conditions of
discontinuity in the electric field drive, specifically when the field
alternates its
sign~\cite{dunlap1986dynamic,Dignam2002conditions,dynamic2002domachuk,dynamic2004wan,quasi2009joushaghani,
  generalized2012joushaghani}.

However, the respective hopping strengths vanish at other points also
individually which results in suppressed transport on tuning the drive
parameters away from EDL conditions. For example, for $p=2,4,6,8,...$,
$\mathcal{J}_{p}^{\text{eff}}$ vanishes when $F/\omega=2m+1, m\in
\mathbb{Z}$, for the specific hopping-range $p$. Similarly, if $F/\omega=r/s$, where $r,s\in \mathbb{Z}$,
$\mathcal{J}_{p}^{\text{eff}}$ will vanish for multiples of $2s$.

Next, we consider the case of disordered hopping ($J=0, \ u_{ij}\neq
0$) in the presence of time-periodic electric field drive, that is,
the periodically driven counterpart of the PLRBM
model~\cite{transition1996Mirlin,anomalous2001cuevas,fluctuations2000evers}. To
understand the driven PLRBM model, we first analyse the
nearest-neighbor counterpart of Eq.~\eqref{eqn:eq1a}, and perform the
Magnus expansion to evaluate the expression for the effective
Hamiltonian,

\begin{eqnarray}
H_{A/B}= -\sum_{n}J_{n}\left(c_{n}^{\dagger}c_{n+1}+h.c.\right)\pm F\sum_{n}n c_{n}^{\dagger}c_{n},
\end{eqnarray}
where $J_{n}\in[-1, 1]$ is the disordered hopping strength drawn from
a uniform distribution. Again, using the BCH formalism~\cite{Hall2015},
we find the effective Hamiltonian given by
\begin{eqnarray}
\label{Eq.Hp}
H_{\text{eff}}&=& -\sum_{n}J_{n}\left(\frac{\sin (FT/4)}{FT/4}\right)\left(c_{n}^{\dagger}c_{n+1}e^{-iFT/4}+h.c.\right)\nonumber\\
&&+H_{1},
\end{eqnarray}
where $J_{n}\in [-W, W],$ and $H_{1}$ contains the higher-order terms
in time-period $T$. In the high-frequency limit, $H_1$ can be
ignored. Thus, electric-field driving again renormalizes the hopping
strength, where tuning the drive-parameters at DL points suppresses
the hopping.

The same analysis can be extended to the disordered model with long
range hopping, however, the effective Hamiltonian will have more
complicated terms (Appendix~\ref{A:1Effective_Hamiltonian}). The
effective Hamiltonian in this case can be expressed as
\begin{eqnarray}
\label{HF_lrd}
H_{\text{eff}}&=& H_{0}+H_{1},\\
\label{HF_0}
H_{0}&=& -\sum_{p=1}^{L-1}\mathcal{J}_{\text{eff}}^{[p]}\left(\hat{K}_{p}e^{-piFT/4}+\hat{K}^{\dagger}_{p}e^{piFT/4}\right),\\
\label{Jff}
\mathcal{J}_{\text{eff}}^{[p]}&=&  \frac{J_{ij}\sin
  (pFT/4)}{p^{\alpha}(pFT/4)},
\end{eqnarray}
where $\mathcal{J}_{\text{eff}}^{[p]}$ is the renormalized hopping strength,
and $H_{1}$ is again higher-order corrections to the effective
Hamiltonian. However, the analysis of Eq.~\eqref{HF_lrd} shows that at
DL points $F=2m\omega$, the zeroth order term $H_{0}$ tends to vanish,
and static and dynamical properties are governed by correction terms
contained in $H_{1}$. We start our numerical analysis with the
diagonalization of the Floquet operator, and obtain Floquet
eigenstates and the quasienergy spectrum.

\subsection{Properties of the Floquet operator}
\begin{figure*}[ht]
  \hspace{-7.7ex}\includegraphics[scale=0.43]{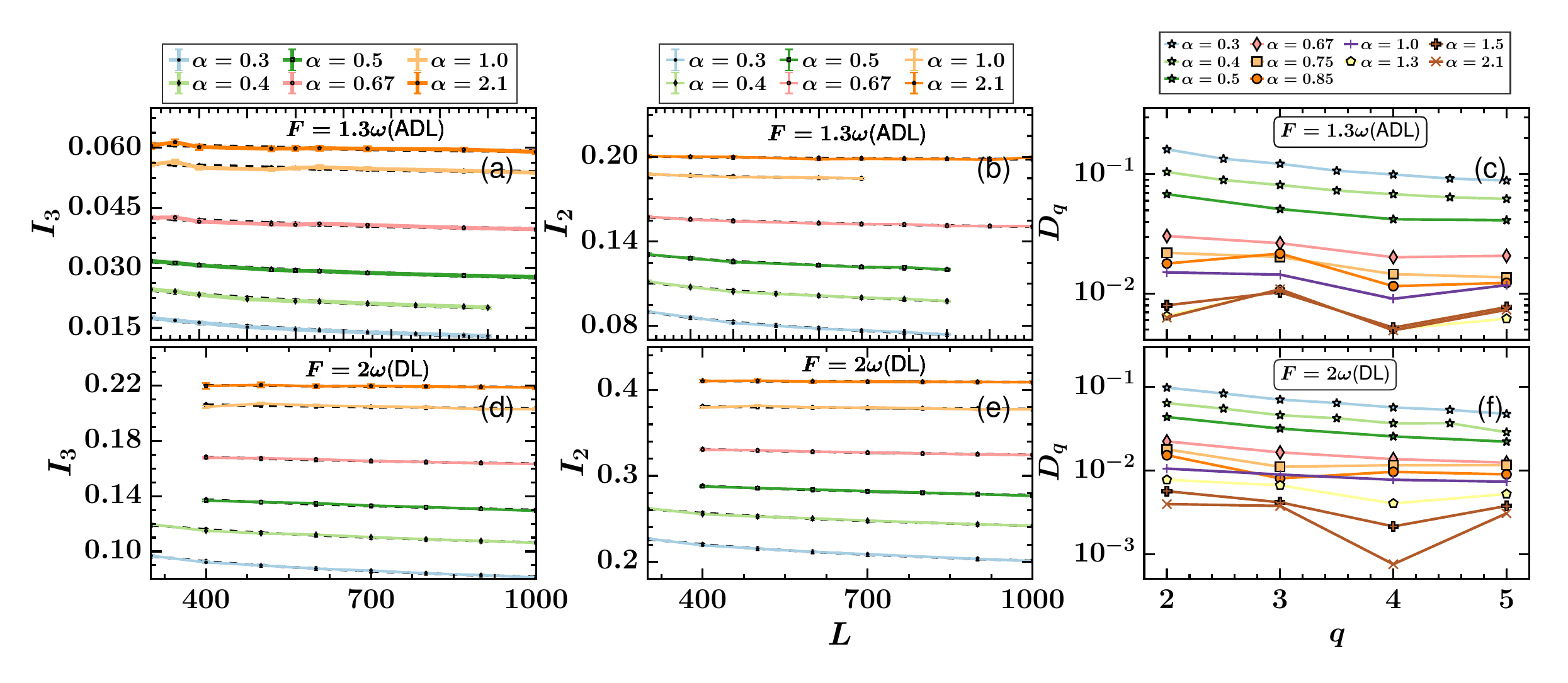}
  \caption{Generalized IPR and fractal dimension $D_{q}$ for
    periodically driven PLRBM model. (a,b) $I_{3}$ and $I_{2}$ vs
    system-size $L$ for drive-parameter $F=1.3\omega$. Black dashed
    line shows power-law fit $(I_{q}\sim AL^{-\tau_{q}})$ to
    characterize the features of Floquet-eigenstates. (d,e) $I_{3}$
    and $I_{2}$ vs system-size $L$ for drive-parameters
    $F=2\omega$. We show the power-law fitting,  $I_{q}\sim
    AL^{-\tau_{q}}$. (c,f) Fractal dimension $D_{q}$ vs $q$ for
    $F=1.3\omega$, and $F=2\omega$, respectively for different
    long-range exponents $\alpha$. The other parameter is
    driving-frequency $\omega=2$. The presented data is averaged over
    $100$ disorder samples. }
  \label{fig:fig1b}
\end{figure*}
To study the static properties of the periodically driven PLRBM model,
we focus on quasienergy eigenvalues and Floquet eigenstates of the
effective Hamiltonian. The quasienergy spectrum plotted in Fig.~\eqref{fig:QE} shows that the gap in the quasienergy spectum is minimum at $F=2m\omega $ i.e. at the zeros of $\mathcal{J}^{[p]}_{\text{eff}}$. However the gap does not vanish compltetely at $F=2m\omega$ due to the presence of correction terms in $H_{1}$ (Eq.~\eqref{Eq.Hp}).
Next, we study the mean of the level spacing
ratio $\langle r\rangle$ between adjacent gaps $\delta$ in the
quasienergy spectrum defined
as~\cite{distribution2013atas,localization2007oganesyan}
\begin{eqnarray}
\label{r_avg}
\langle r\rangle &=&\left\langle \frac{\text{min}\left(\delta_{j}, \delta_{j+1}\right)}{\text{max}\left(\delta_{j}, \delta_{j+1}\right)}\right\rangle ,
\end{eqnarray}
where $\delta_{j}=\epsilon_{j+1}-\epsilon_{j}$, and $\epsilon_{j}$ is
the quasienergy eigenvalue. In the delocalized phase, the average gap
ratio $\langle r\rangle$ approaches the circular orthogonal ensemble
(COE) value, $\langle r\rangle=0.529$, and in the localized phase, the
average gap ratio approaches the value $\langle
r\rangle=0.386$~\cite{localization2007oganesyan,distribution2013atas,ponte2015many}
which is consistent with gaps obeying the Poisson distribution. In
Fig.~\eqref{fig:fig1a}(a), the average gap ratio $\langle r\rangle$ is
plotted as a function of the long-range exponent $\alpha$ for drive
parameters tuned at ADL ($F=1.3\omega$) and DL ($F=2\omega$) points of
the clean limit. For $\alpha > 1$, the value of $\langle r \rangle\approx 0.386$ indicates that the periodically
  driven PLRBM model remains localized for both the field strengths, $F = 1.3\omega$ and $F = 2\omega$. For $\alpha
  < 1$, $\langle r \rangle$ lies between the COE value ($\langle r
  \rangle = 0.529$) and the Poisson value ($\langle r \rangle =
  0.386$), suggesting the emergence of an intermediate phase where the
  static system exhibits delocalized behavior.

The inset in Fig.~\eqref{fig:fig1a}(a) shows the variation of $\langle r \rangle$ with $F/\omega$, highlighting distinct phases for $\alpha < 1$ and $\alpha > 1$. The sudden dips at $F/\omega = 2m$ result from the simultaneous vanishing of all the hopping components of $H_0$ (Eq.~\eqref{HF_0}). However, correction terms present in $H_1$ cause deviations from the Poisson statistics. Near the dynamical localization (DL) points ($F = (2m \pm \delta)\omega$), the behavior is governed by:  
\begin{equation}
\sin(p\pi \pm p(\pi/2)\delta) = \pm \sin(p(\pi/2)\delta),
\end{equation}
where the RHS vanishes for larger $p$. Smaller $\delta$ values result in a comparatively long-range hopping model, which explains the observed sharp peaks near the DL points.  

The sudden dips at $F = 2m\omega$ (DL) (inset:
Fig.~\eqref{fig:fig1a}(a)) are further analyzed in
Fig.~\eqref{fig:fig1a}(b,c) using the hopping amplitude of the
zeroth-order effective Hamiltonian, $\mathcal{J}_{\text{eff}}^{[p]}$
(Eq.~\eqref{Jff}). To quantify this, we compute:
\begin{eqnarray}
\label{I_hp}
\eta = \frac{\sum_p \left( \mathcal{J}_{\text{eff}}^{[p]} \right)^4}{\left( \sum_p \left( \mathcal{J}_{\text{eff}}^{[p]} \right)^2 \right)^2}, \quad 
\langle \mathcal{J}_{\text{eff}}\rangle= \frac{1}{L} \sum_p \left( \mathcal{J}_{\text{eff}}^{[p]} \right),
\end{eqnarray}
$\eta$ is a measure of the effective number of bonds which have a
significant hopping strength, while $\langle
\mathcal{J}_{\text{eff}}\rangle$ is the average of all the hoppings
from a given site.  In Fig.~\eqref{fig:fig1a}(b), $\eta$ exhibits dips
at the DL points, indicating negligible contributions from hopping,
which is further confirmed by the negligible values of
$\langle \mathcal{J}_{\text{eff}}\rangle$ (Fig.~\eqref{fig:fig1a}(c)). As $\alpha$ increases
(from $\alpha = 0.3$ to $\alpha = 10$), contributions from certain
sites in the one-dimensional system become more significant, leading
to higher dip values at the DL points.  The peaks of $\eta$ at $F =
(2m+1)\omega$ in Fig.~\eqref{fig:fig1a}(b) correspond to the presence
of non-zero hopping components along with zero hopping components at
$p = (2m+1)$ (Eq.~\eqref{I_hp}). However, as $\alpha$ increases, the
effective number of non-zero hopping components decreases, leading to
higher peak values of $\eta$. This feature is also captured by
$\langle \mathcal{J}_{\text{eff}}\rangle$, which shows dips at $F = (2m+1)\omega$
(Fig.~\eqref{fig:fig1a}(c)).

We next study the generalized average inverse participation ratio
(IPR) of the Floquet eigenstates $I_{q}$, and fractal dimension
$D_{q}$~\cite{multifractality2018sthitadhi},
\begin{eqnarray}
I_{q}= \sum_{i=1}^{N} |\psi_{i}(l)|^{2q}\sim L^{-\tau_{q}},\quad \quad D_{q}=\frac{\tau_{q}}{q-1},
\end{eqnarray}
where $|\psi(l)\rangle$ is the $l^{\text{th}}$ normalized Floquet eigenstate,
and can be expanded in Wannier-basis $|i\rangle $ as
$|\psi(l)\rangle=\sum_{i=1}^{N}\psi_{i}(l)$, and $D_{q}$ denotes the
fractal dimension. The scaling of $I_{2}$ with system size $L$ helps
characterize the phases: localized phase ($I_{2}\sim L^{0}$),
delocalized phase ($I_{2}\sim L^{-1}$), and multifractal phase
($I_{2}\sim L^{-\tau_{q}}, 0<\tau_{q} <1 $). Furthermore, the algebraic
scaling of generalized inverse participation ratio ($ I_{q}$) with the
system size $L$ yields the fractal dimension $D_{q}$ which carries
complete information essential to characterize the
multifractality. For the localized to delocalized phase, $D_{q}$
varies from $0$ to $1$, and intermediate values of $D_{q}$ (with a
non-trivial dependence on $q$) implies the multifractality of the
Floquet eigenstates. However, the independence of $D_{q}$ on $q$ implies the fractal behavior of Floquet eigenstates.
\begin{figure*}[t]
	\includegraphics[width=0.49\textwidth]{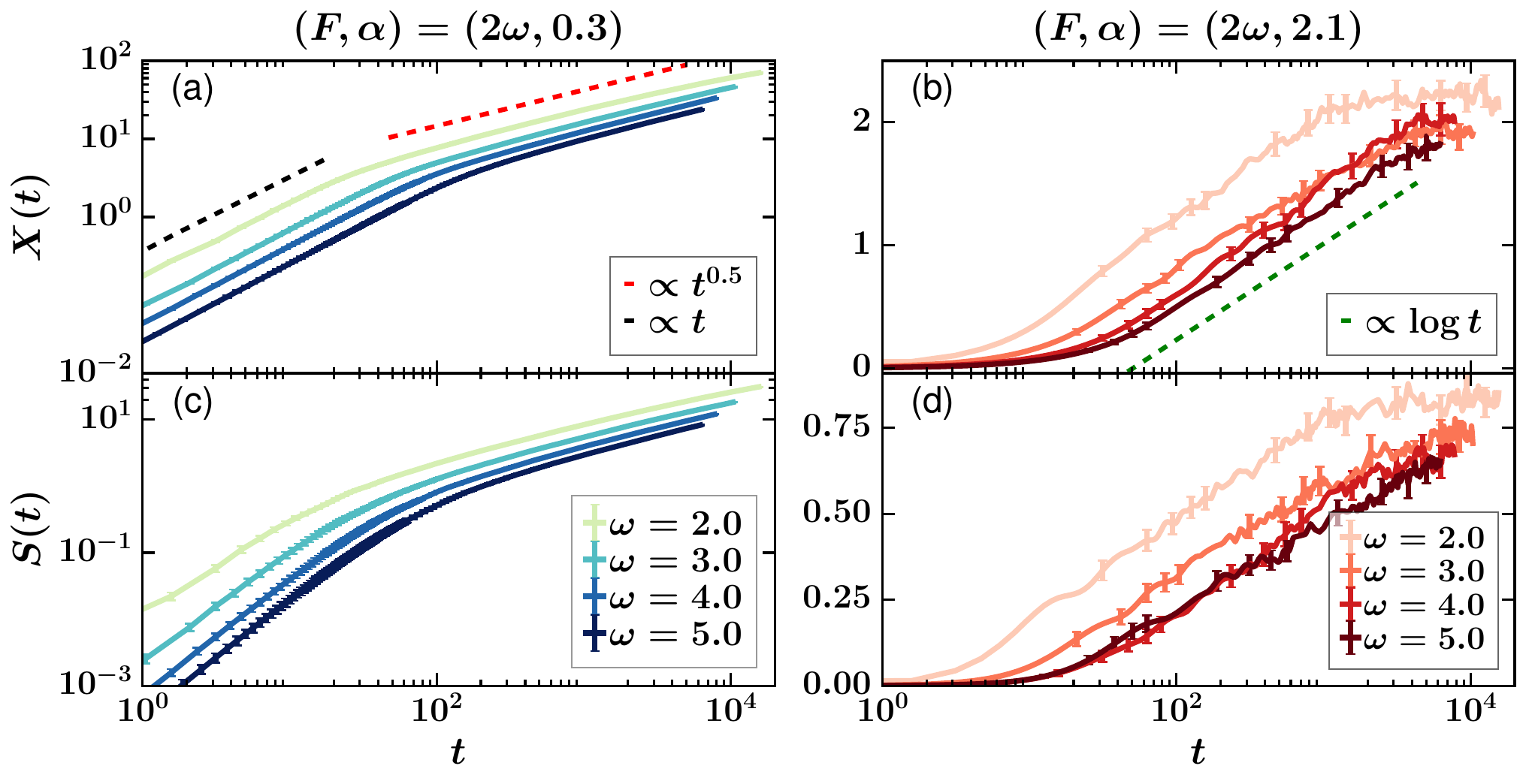}
	\includegraphics[width=0.49\textwidth]{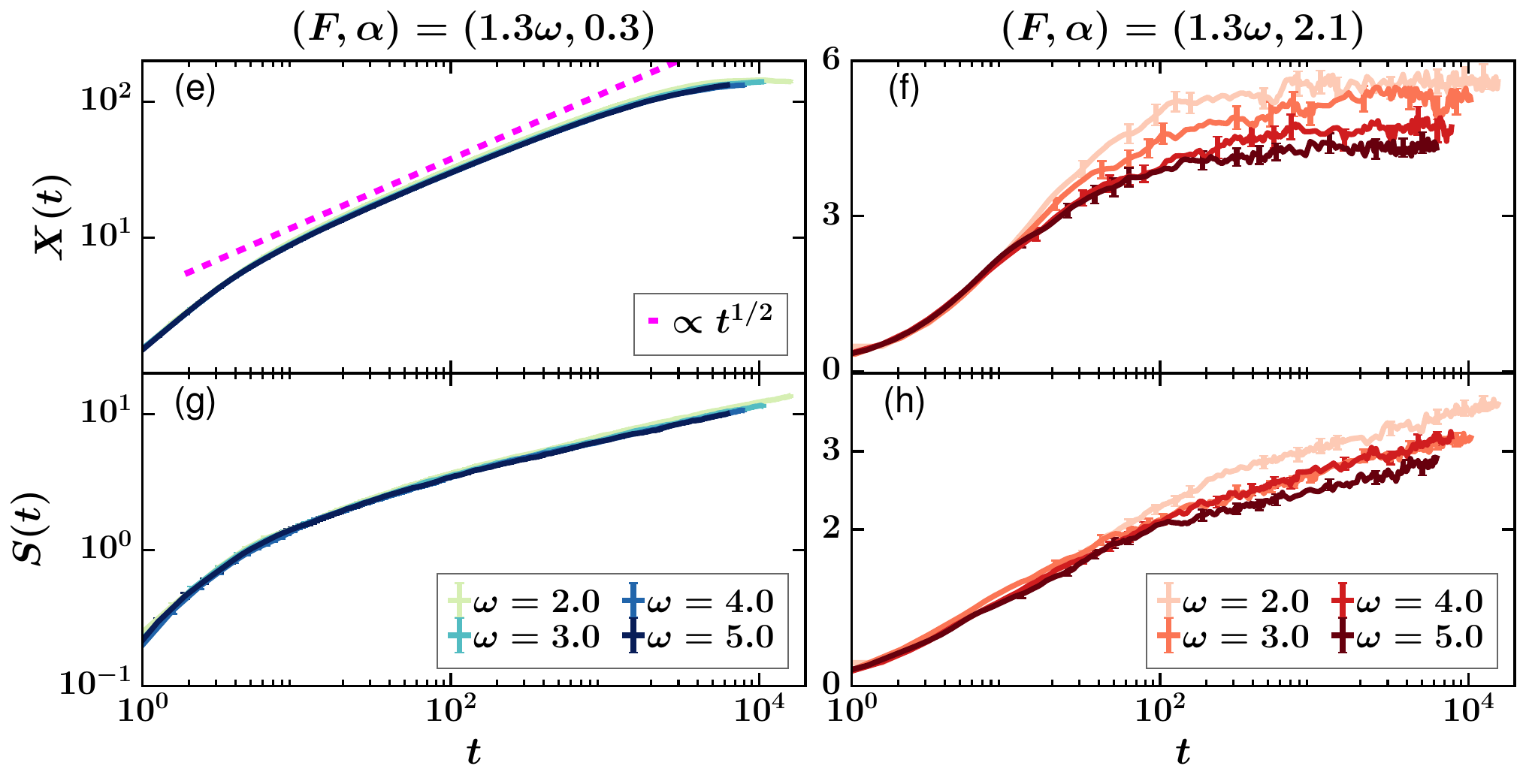}
	\caption{(a-d). Dynamics of periodically driven system tuned at DL points $F=2\omega$ at frequencies $\omega=2,3,4,5$ for long-range exponent $\alpha=0.3, 2.1$. (a,c) The dynamics of RMSD $X(t)$ and entanglement entropy $S(t)$ for $\alpha=0.3$. The dashed lines in (a) show two transport regimes: (i). transient ballistic transport $X(t)\propto t$ shown in black color, (ii). diffusive transport $X(t)\propto t^{0.5}$ shown in red color. (b,d) The dynamics of RMSD $X(t)$ and entanglement entropy $S(t)$ for $\alpha=2.1$. The dashed line shows logarithmic transport, $X(t)\propto \log t$. (e-h). Dynamics of periodically driven system at ADL points, $F=1.3\omega$ at frequencies $\omega=2,3,4,5$ for long-range exponent $\alpha=0.3, 2.1$. (e,g). The dynamics of RMSD $X(t)$ and entanglement entropy $S(t)$ for $\alpha=0.3$. The dashed pink line shows diffusive transport, $X(t)\propto t^{0.5}$. (f,h). The dynamics of RMSD $X(t)$ and entanglement entropy $S(t)$ for $\alpha=2.1$. The system-size considered is $L=1024$.}
	\label{fig:fig2}
\end{figure*}
To quantify the localization and delocalization properties of the
Floquet eigenstates, we plot $I_{q}$ vs $L$ in Fig.~\eqref{fig:fig1b}
for the field strength $F=1.3\omega$ (Fig.~\eqref{fig:fig1b}(a,b)),
and $F=2\omega$ (Fig.~\eqref{fig:fig1b}(d,e)). We find that the
variation in $I_{q}$ with $L$ decreases with increase in the
long-range exponent $\alpha$. Fig.~\eqref{fig:fig1b} clearly
demonstrates that $I_{q}$ varies with $L$ for $\alpha<1$, and becomes
almost constant for $\alpha>1$. This shows that the periodically
driven PLRBM model remains localized for $\alpha>1$ irrespective of
the driving parameters.  Furthermore, to quantify the degree of
localization and delocalization, we perform a power-law fit
$(I_{q}=AL^{-\tau_{q}})$ of $I_{q}$ vs $L$ for $q=2,3$ in
Fig.~\eqref{fig:fig1b}(a,b,d,e) (shown in black dashed lines), and
extract the the coeffecients for fractal dimension $D_{q}$ in
Fig.~\eqref{fig:fig1b}(c,f). The study of $D_{q}$ as a function of $q$ shows that for $\alpha<1$,
the fractal dimension lies in the range $0<D_{q}<1$. For $\alpha=0.3$ and $0.4$, $D_{q}$ exhibits weak dependence on $q$, indicating the presence of weak multifractality as shown in Fig.~\eqref{fig:fig1b}(c,f). However, as 
$\alpha $ increases while remaining on the delocalized side of the PLRBM model, $D_q$
remains largely constant with varying $q$ (Fig.~\eqref{fig:fig1b}(c,f)). This indicates that the driven
system exhibits fractal behavior as shown for $\alpha=0.5,0.67,0.75,$ and $0.85$ in Fig.~\eqref{fig:fig1b}(c,f). On the other hand, for $\alpha>1$,
the fractal dimension $D_{q}\approx 0$ (Fig.~\eqref{fig:fig1b}(c,f)),
and hence indicates the localization of the Floquet eigenstates.
\begin{figure}[t]
\centering
\hspace{-2ex}\includegraphics[width=0.45\textwidth]{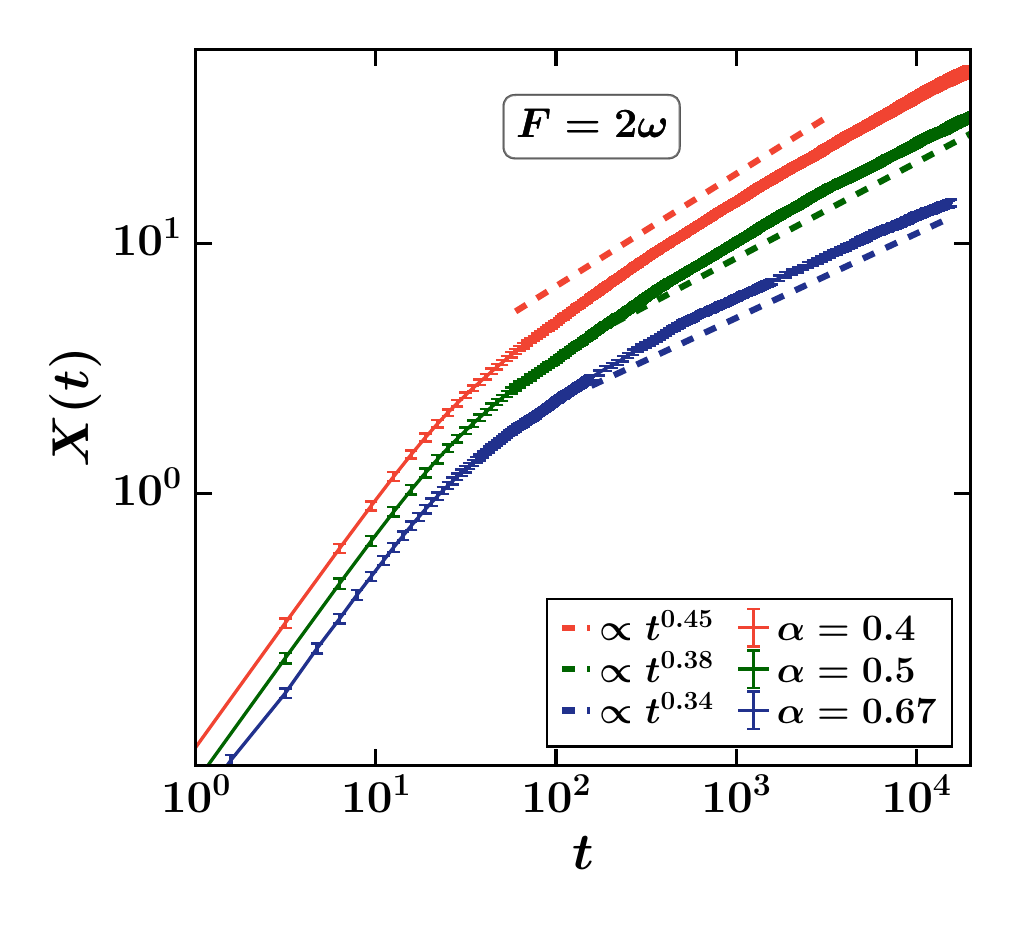}
\caption{Dynamics of root-mean-squared-width $X(t)$ for periodically driven system tuned at DL points $F=2\omega$ for different long-range exponents $\alpha$. The dashed lines correspond to power-law fit $X(t)\propto t^{\beta}$ for the curves plotted in the same color. The other parameters are system-size $L=1024,$ and driving-frequency $\omega=2$.}
\label{fig:fig2b}
\end{figure}
The emergence of weak multifractal and fractal behavior of the eigenstates for $\alpha<1$
can be understood with the aid of an analytical expression for the
effective Hamiltonian $H_{\text{eff}}$ (Eq.~\eqref{HF_lrd}). For the
tuning of the drive parameters at DL points $(F=2m\omega)$, the zeroth
order term $H_{0}$ in the effective Hamiltonian (Eq.~\eqref{HF_0}) tends to
vanish as the renormalized hoppings $\mathcal{J}_{\text{eff}}^{[p]}$
vanish for all the ranges of hopping $(p)$, and the effective
Hamiltonian is governed by the higher order correction terms contained
in $H_{1}$. In contrast to the DL case, the zeroth order term $H_{0}$
survives on tuning the drive-parameters at ADL points. However, some
hopping components of the renormalized hopping strength
$\mathcal{J}_{\text{eff}}^{[p]}$ tends to vanish given the condition
that for $F\neq 2m\omega$, there exist $p^{\prime}$s where
$\mathcal{J}_{\text{eff}}^{[p]}$ vanishes.  This results in
\textit{drive-induced shortening of the range of hopping} in the
effective Hamiltonian. In other words, electric-field periodic drive
effectively suppresses the range of hopping even at ADL points, and
transitions the system from the delocalized phase ($\alpha<1$) to near
the transition limit where the undriven Hamiltonian is known to show
multifractality~\cite{transition1996Mirlin}. As soon as the long-range
exponent $\alpha$ enters into the localized regime of the undriven
model, the interplay of long-range hopping and external periodic
driving gives rise to the localized phase irrespective of the
drive-parameters as characterized by higher saturation values of
$I_{q}$, and smaller values of $D_{q}$ for $\alpha>1$.
\begin{figure*}[t]
\centering
	\hspace{-6ex}\includegraphics[width=0.4\textwidth]{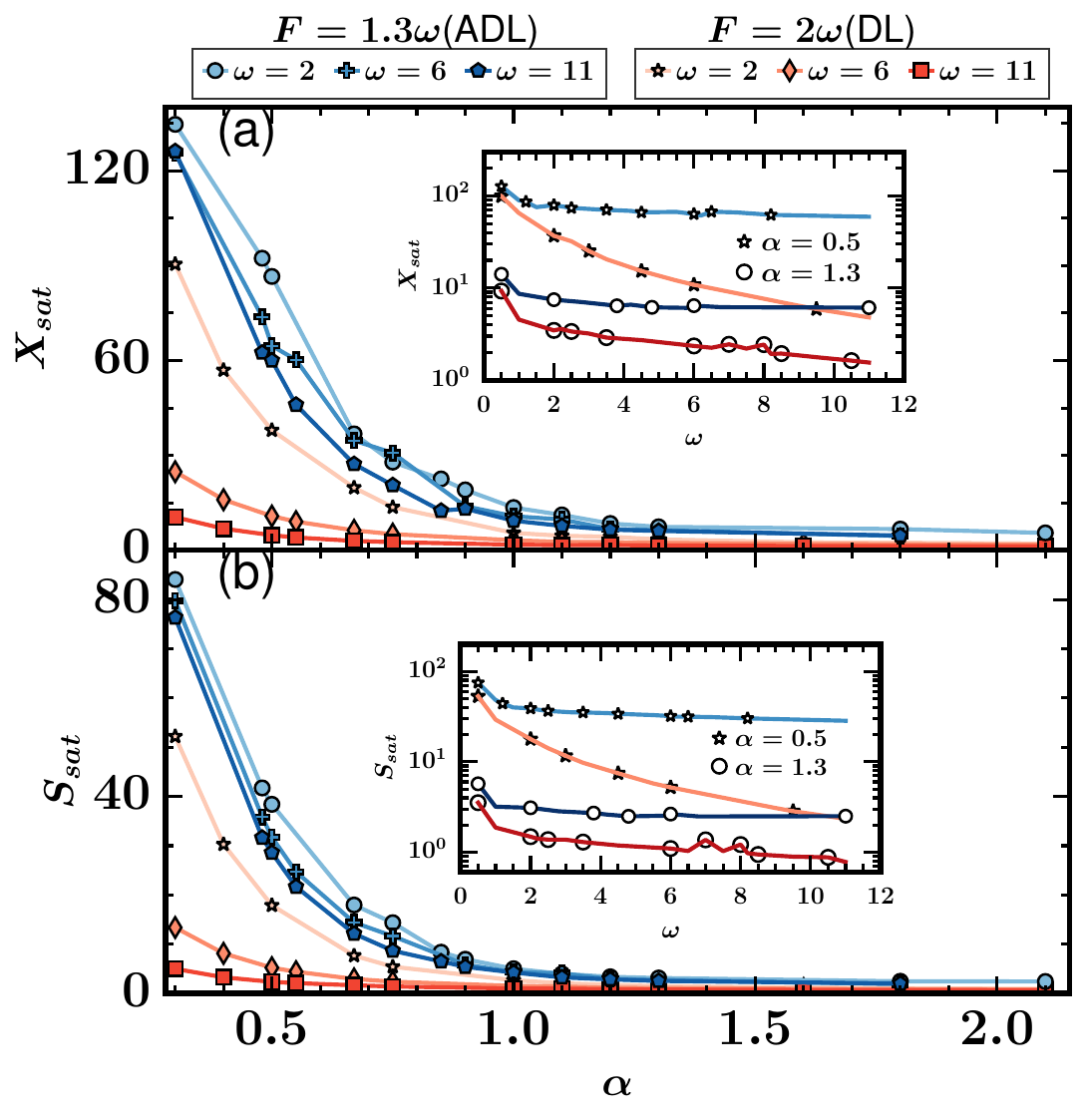}
	\hspace{6ex}\includegraphics[width=0.4\textwidth]{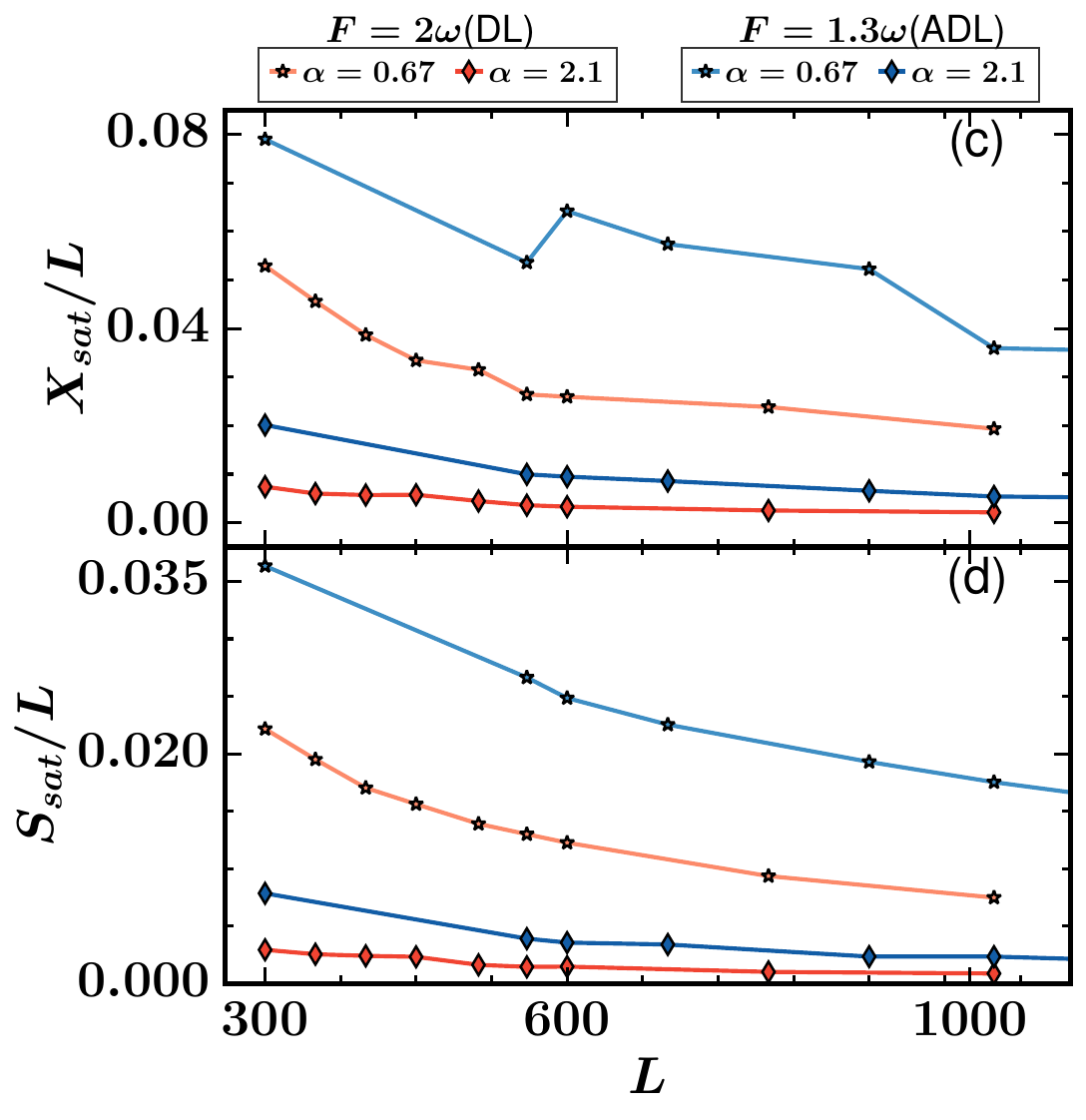}
\caption{(a,b). Saturation values of root-mean squared displacement $X_{sat}$ and entanglement entropy $S_{sat}$ with long-range exponent $\alpha$ for driving frequencies $\omega=2,6,11$. Inset(a,b): $X_{sat}$ and $S_{sat}$ with driving-frequency $\omega$ for long-range exponents $\alpha=0.5, 1.3$. In the plots, the red shades correspond to the drive-parameters tuned at DL points of clean limit, $F=2\omega$, and blue colour plots correspond to the tuning away from DL points of clean limit, $F=1.3\omega$. The other parameters are $L=1024, J=1$. (c,d).The decays of saturation values of root-mean squared displacement $X_{sat}$ and entanglement entropy $S_{sat}$ with system size $L$ for driving frequency $\omega=2$, and long-range exponents $\alpha=0.67,2.1$. We computed the saturation values after $n=10000$ cycles.}
	\label{fig:fig3a}
\end{figure*} 
Moreover, a careful analysis of Fig.~\eqref{fig:fig1b} shows that at
DL points, the IPR increases with the increase in the long-range exponent,
and hints at localization beyond $\alpha=1$, albeit with an extended
localized length compared to the localization in the static
case. The emergence of localization-like behavior
at DL points is in strong agreement with the recent experimental
observations in a periodically driven AAH
system~\cite{dotti2024measuring}. However, our finding of a weakly delocalized fractal phase on the delocalized side of the undriven PLRBM model differs from Ref.~\cite{dotti2024measuring}.

\subsection{Transport properties}

To explore the dynamical phases of the driven PLRBM system, we study
the transport properties with the help of key observables: root
mean-squared displacement(RMSD) ($X(t)$)(Eq.~\eqref{eq:msd}) and
entanglement entropy ($S(t)$)(Eq.~\eqref{Sent}). We analyze the
dynamics for two long-range exponents $\alpha=0.3, 2.1$ at DL
($F=2\omega$) (Fig.~\eqref{fig:fig2}(a-d)) and ADL ($F=1.3\omega$)
(Fig.~\eqref{fig:fig2}(e-h)) points of the clean limit. This
comparison enables us to characterize the dependence of dynamical
features on the long-range exponent $\alpha $, and the
drive-parameters.

The static PLRBM model exhibits ballistic behavior in the delocalized
phase and an absence of transport in the localized phase. In contrast
to the static case, the periodically driven model reveals dynamical
crossover from a transient ballistic regime $(X(t)\propto t)$ to the
diffusive regime $(X(t)\propto t^{0.5})$ for $\alpha<1$ as shown in
Fig.~\eqref{fig:fig2}(a,c) when the drive-parameters are tuned at DL
points, $F=2\omega$. However, on increasing $\alpha$, the dynamics of RMSD $X(t)$ follows subdiffusive transport, $X(t)\propto t^{\beta}$($\beta<0.5$) for $\alpha=0.5, 0.67$ as shown in Fig.~\ref{fig:fig2b}. Hence, the fractal behavior of eigenstates is accompanied by anomalous subdiffusive transport, and is consistent with the studies where multifractality and slow dynamics have been reported together~\cite{ketzmerick1992slow,ketzmerick1997what,ohtsuki1997anomalous,multifractality2018sthitadhi,Duthie2022anomalous}.  On the localized side of the PLRBM model
$(\alpha>1)$, $X(t)$ and $S(t)$ exhibit suppressed logarithmic growth
$(X(t), S(t)\propto \log t)$, followed by asymptotic saturation to
values much lower than those of the infinite temperature state, as
illustrated in Fig.~\eqref{fig:fig2}(b and d) for long-range exponent
$\alpha=2.1$. The suppressed transport emerges as an effective result
of driving a long-range system at DL points where the interplay of
hopping suppression (Eq.~\eqref{HF_0}) and the exponent being tuned to
be short-range $(\alpha>1)$ plays a key role. The
  dependence of the dynamics on the driving frequencies at DL points
  for $\alpha<1$ and $\alpha>1$ (Fig.~\eqref{fig:fig2}(a-d)) can again
  be explained with the help of an effective Hamiltonian
  (Eq.~\eqref{HF_lrd}). At DL points, the effect of $H_{0}$ remains
  valid in the short-time limit where the dynamics exhibits frozen
  behavior as an effect of $\mathcal{J}_{\text{eff}}^{[p]}\rightarrow
  0$ (as shown in Fig.~\eqref{fig:fig1a}(b,c)). However, the dynamics
  is effectively governed by the correction terms $H_{1}$ in the
  long-time limit (Eq.~\eqref{HF_lrd}). Consequently, increasing the
  driving frequency truncates the Magnus expansion at higher orders,
  thereby extending the relaxation time~\cite{mori2016rigorous}.

Next, we discuss the dynamics of $X(t)$ and $S(t)$ at ADL points
$(F=1.3\omega)$ for the long-range system with $\alpha=0.3, 2.1$
(Fig.~\eqref{fig:fig2}(e-h)). On the delocalized side $(\alpha<1)$,
$X(t)$ and $S(t)$ exhibit diffusive transport ($X(t)\propto t^{0.5}$)
accompanied by aysmptotic relaxation to the inifinite temperature
state at $\alpha=0.3$. In contrast, on the localized side $(\alpha>1
)$, $X(t)$ and $S(t)$ exhibit asymptotic saturation to values much
lower than those of the infinite temperature
state. Fig.~\eqref{fig:fig2}(f,h) show that the system shows
localization signatures even in the case of driving at ADL points;
driving at higher frequencies enhances the localization tendency
because of the truncation of the higher-order long-range terms in the
Magnus expansion. However, in contrast to the dynamics at
  DL points, the dynamics of $X(t)$ and $S(t)$ does not exhibit
  frequency dependence for $\alpha=0.3, 2.1$ (
  (Fig.~\eqref{fig:fig2}(e-h)) ). Unlike the DL case, the dynamics is
  effectively governed by both $H_{0}$ and $H_{1}$. The form of
  $H_{0}$ represents the PLRBM Hamiltonian with renormalized hopping
  magnitude. On increasing the driving-frequency,
  $\mathcal{J}_{\text{eff}}^{[p]}\rightarrow J_{ij}$, and $H_{0}$
  becomes frequency independent. Hence, in the short time-limit, the
  dynamics is dominated by $H_{0}$ and exhibits
  frequency-independence, however in the long-time limit, the effect
  of $H_{1}$ becomes significant and we do observe some frequency
  dependence (Fig.~\eqref{fig:fig2}(f,h)).

To show the comparison between the dynamics of the system driven at DL
and ADL points of the clean limit, we compute saturation values of
RMSD $X_{sat}$ and entanglement-entropy $S_{sat}$
(Fig.~\eqref{fig:fig3a} at driving cycle $n=10000$ for the
driving-parameters tuned at $F=2\omega$ (red colour plots) and
$F=1.3\omega$ (blue colour plots). Fig.~\eqref{fig:fig3a}(a) and
Fig.~\eqref{fig:fig3a}(b) show that $X_{sat}$ and $S_{sat}$ decay with
the long-range exponent $\alpha$. The saturation of the quantities
remains higher on tuning the drive-parameters at $F=1.3\omega$
compared to the points tuned at $F=2\omega$. However, for both the
tunings, saturation values exhibit significant suppression for
$\alpha>1$. The inset plots (Fig~\eqref{fig:fig3a}(a,b)) show that
$X_{sat}$ and $S_{sat}$ exhibit frequency dependence on the
driving-frequency $\omega$ for $F=2\omega$(red color plots). On the
other hand, the saturation values do not show significant dependence
on driving-frequency $\omega$ for $F=1.3\omega$.

Thus, the combined study of the time-periodic system at DL points
(Fig.~\eqref{fig:fig2}(c,d)) and ADL points
(Fig.~\eqref{fig:fig2}(f,h)) suggests that driving the system at
higher frequency truncates the Floquet Hamiltonian asymptotically to
$H_{0}$ which is the PLRBM model with the renormalized hopping strength
$\mathcal{J}_{\text{eff}}^{[p]}$ given by (Appendix~\ref{A:1Effective_Hamiltonian}):
\begin{eqnarray}
\label{HF_l}
H_{\text{eff}}\approx H_{0} &=& -\sum_{p=1}^{L-1}\mathcal{J}_{\text{eff}}^{[p]}\left(\hat{K}_{p}e^{-piFT/4}+\hat{K}^{\dagger}_{p}e^{piFT/4}\right),
\nonumber\\
\mathcal{J}_{\text{eff}}^{[p]} &=& \frac{J_{p}\sin (pFT/4)}{p^{\alpha}(pFT/4)}.
\end{eqnarray}
\begin{figure}[t]
\centering
\hspace{-2ex}\includegraphics[width=0.47\textwidth]{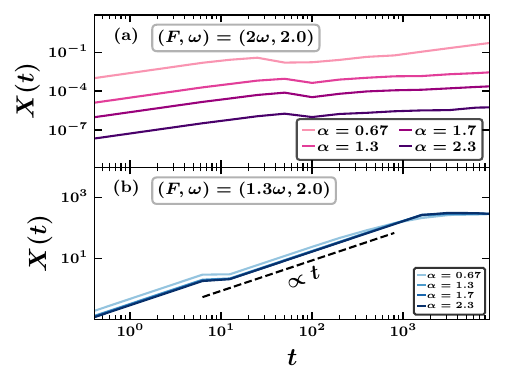}
\caption{Root mean squared displacement for Thue-Morse driven clean
  long-range hopping system at dynamical localization point $F=2\omega
  $ (top) and away from dynamical localization point $F=1.3\omega$
  (bottom) for various values of the long-range exponent. Dashed line
  show the power law fitting ($X(t)\propto t$) for the ballistic
  transport. The other parameters are $J=1.0, L=1000$. }
\label{fig:fig5a}
\end{figure}

This results in weak delocalization (weak multifractal and fractal) both at DL and ADL
points for $\alpha<1$. In contrast, driving the long-range system with
$\alpha>1$ yields localization with the ADL points exhibiting
localization with larger localization length. At DL points, however,
hopping suppression results in reduced saturation values of $X(t)$ and
$S(t)$, along with an increased average inverse participation ratio
(IPR) for $\alpha>1$. This suggests that the system remains localized
akin to the many-body localized phase where the entanglement also
grows logarithmically ($S(t)\propto \log t$). Such a logarithmic slow
transport could arise from the spreading of the quasienergy band at
the dynamical localization point. Thus, our study of the transport
properties supports the results obtained from the analysis of the
Floquet Hamiltonian.

\begin{figure*}[t]
\centering
	\hspace{-2ex}\includegraphics[width=1\textwidth]{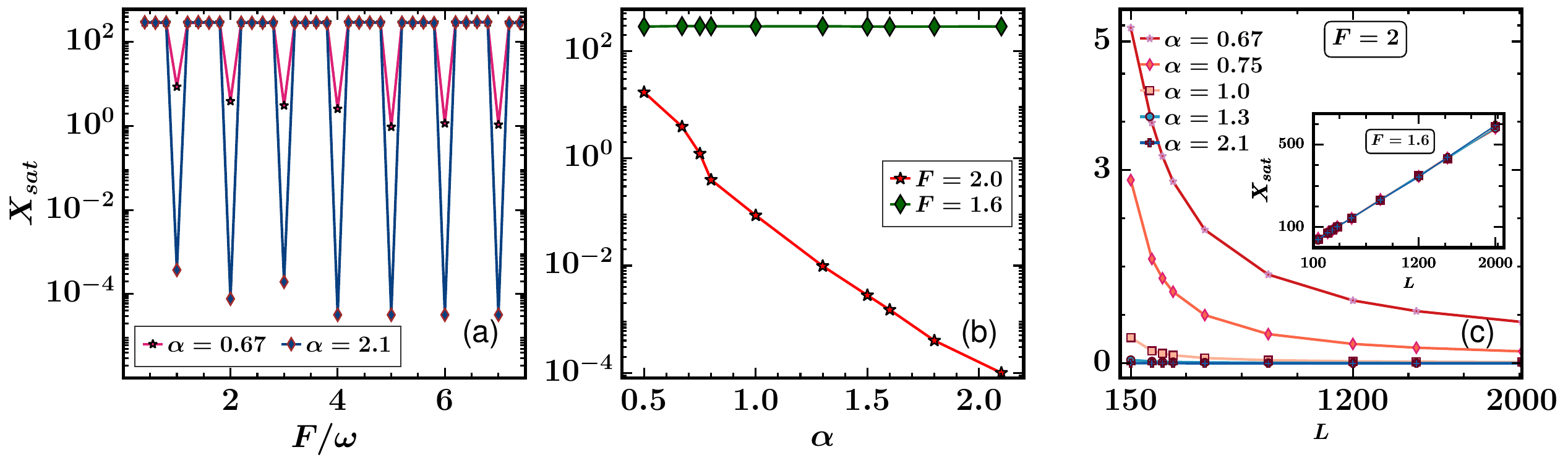}
\caption{(a) Oscillatory behavior of saturation values of root-mean squared displacement $X_{sat}$ with varying values of drive-paramters $F/\omega$ for long-range exponent $\alpha$. The other parameters are $L=1000, J=1, \omega=2$. (b) $X_{sat}$ with $\alpha$ for the drive-parameters $F=2, 1.6$, and frequency $\omega=1$. (c) $X_{sat}$ with system-size $L$ for different values of $\alpha$ and drive-parameters $F=2,\omega=1$ (main plot), and $F=1.6,\omega=1$ (inset plot). We compute saturation values $X_{sat}$ at Thue-Morse cycle $n=15$.}
	\label{fig:fig5b}
\end{figure*}

\section{Aperiodic Thue-Morse driving}
\label{Aperiodic Thue-Morse Driving}
We now consider the case where the time-dependent field is not
periodic and is taken from a Thue-Morse
sequence~\cite{tiwari2024dynamical}.  We look at both clean long-range
hopping as well as disordered long-range coupling.

\subsection{Clean long-range hopping} 
We first consider the clean long-range hopping limit
(Eq.~\eqref{eqn:eq1a}) with $u_{ij}=0$ and explore if it is possible
to observe dynamical localization even for the aperiodic Thue-Morse
driving protocol and for arbitrary long-range parameter $\alpha$.  To
obtain the conditions for the dynamical localization, we perform the
high-frequency expansion similar to the periodically driven case but
with the recurrence relation for the Thue-Morse sequence defined in
Eq.~\eqref{TMS_U}. We first evaluate the unitary operators $U_{1}$ and
$\tilde{U}_{1}$ using Baker-Campbell-Hausdorff formula~\cite{Hall2015}
as

\begin{eqnarray}
	U_{1}&=& U_{B}U_{A}=\exp(-iTH_{BA}^{\text{F}}),\\
	H_{BA}^{\text{F}}&=& -\sum_{p>0}J_{p}^{\text{F}}\left(\hat{K}e^{-iFT/2}+\hat{K}^{\dagger}e^{iFT/2}\right),\\
	\tilde{U}_{1}&=& U_{A}U_{B}=\exp(-iTH_{AB}^{\text{F}}),\\
	H_{AB}^{\text{F}}&=& -\sum_{p>0}J_{p}^{\text{F}}\left(\hat{K}e^{iFT/2}+\hat{K}^{\dagger}e^{-iFT/2}\right),
\end{eqnarray}
where, $J_{p}^{\text{F}}= J_{p}\frac{\sin
  \left(pFT/2\right)}{\left(pFT/2\right)},$ and
$J_{p}=\frac{J}{p^{\alpha}}$. It is evident from the recurrence
relation (Eq.~\eqref{TMS_U}) that the Thue- Morse sequence always
yields pairs of $U_{A}$ and $U_{B}$, with the total number of these
operators being an exact even power. We first start at Thue-Morse
level $2$ with $2^{m=2}$ pulses: $A, B, B, A$. The time-evolution
operator can be written in a simplified form as
\begin{eqnarray}
	U(m=2)&=& U_{A}U_{B}U_{B}U_{A}=\exp(-2^{2}iTH_{\text{eff}}),
\end{eqnarray}
where $H_{\text{eff}}$ is the effective Hamiltonian defined as
\begin{equation}
	\label{Hf_T}
	H_{\text{eff}}= -\sum_{p}J_{p}^{\text{eff}}\left(\hat{K}_{p}+\hat{K}^{\dagger}_{p}\right),J_{p}^{\text{eff}}= J_{p}\frac{\sin \left(pFT\right)}{\left(pFT\right)}.
\end{equation}
Here, $J_{p}^{\text{eff}}$ is the effective hopping strength which vanishes for the dynamical localization condition $F=n\omega (FT=n\pi)$. Furthermore, the construction of the time-evolution operator $U(N=2^{m})$ for the Thue-Morse sequence leads to the generalized form of the unitary operator:
\begin{eqnarray}
\label{UF_T}
	U(N=2^{m})&=&\exp(-2^{m}iTH_{\text{eff}}).
\end{eqnarray} 
The expression for the effective Hamiltonian (Eq.~\ref{Hf_T}) suggests
that the drive renormalizes the hopping parameter $J_{p}$ to
$J_{p}^{\text{eff}}$ and hence, at the zeros of $J_{p}^{\text{eff}}$,
$F=n\omega$, one can observe the phenomenon of \textit{exact dynamical
  localization} similar to the case of square-wave driving. At these
special points, the transport is suppressed completely
(Fig.~\eqref{fig:fig5a}(a)). On the other hand, away from these
parameters, the system behaves as a tight-binding chain with
renormalized hopping. Thus, the system features ballistic transport,
$X(t)\propto t$, on tuning the drive parameters away from the
dynamical localization point(Fig.~\eqref{fig:fig5a}(b)).

To verify our analytical results (Eq.~\eqref{Hf_T},\eqref{UF_T}), we
present numerical analysis and plot the dynamics of the mean squared
displacement $X(t)$ in Fig.~\ref{fig:fig5a} for a range of long-range
hopping parameters $\alpha=0.67,1.3,1.7,2.3$. We fix the driving
parameters to correspond to a dynamical localization point
($F/\omega=2$) and also to an away from the dynamical localization
point ($F/\omega=1.3$). It can be seen from Fig.~\ref{fig:fig5a}(a)
that for the parameters tuned at dynamical localization, $X(t)\approx
0$, indicating the absence of transport, and confirming the robustness
of the dynamical localization for arbitrary long-range hopping.

We elaborate on our numerical computation shown in
Fig.~\eqref{fig:fig5a}, and analyze the saturation values of $X(t)$
obtained at Thue-Morse cycle $n=15$ ($X_{sat}$) with the system
parameters in Fig.~\eqref{fig:fig5b}. Fig.~\eqref{fig:fig5b}(a)
illustrates the oscillatory behavior of $X_{sat}$ with $F/\omega$
where dips lie at $F=n\omega$, $n\in \mathbb{Z}$, and constant
saturation values at $F\neq n\omega$ for long-range exponent
$\alpha=0.67,2.1$. 
\begin{figure*}[ht!]
	\hspace{-3ex}\includegraphics[width=0.62\textwidth]{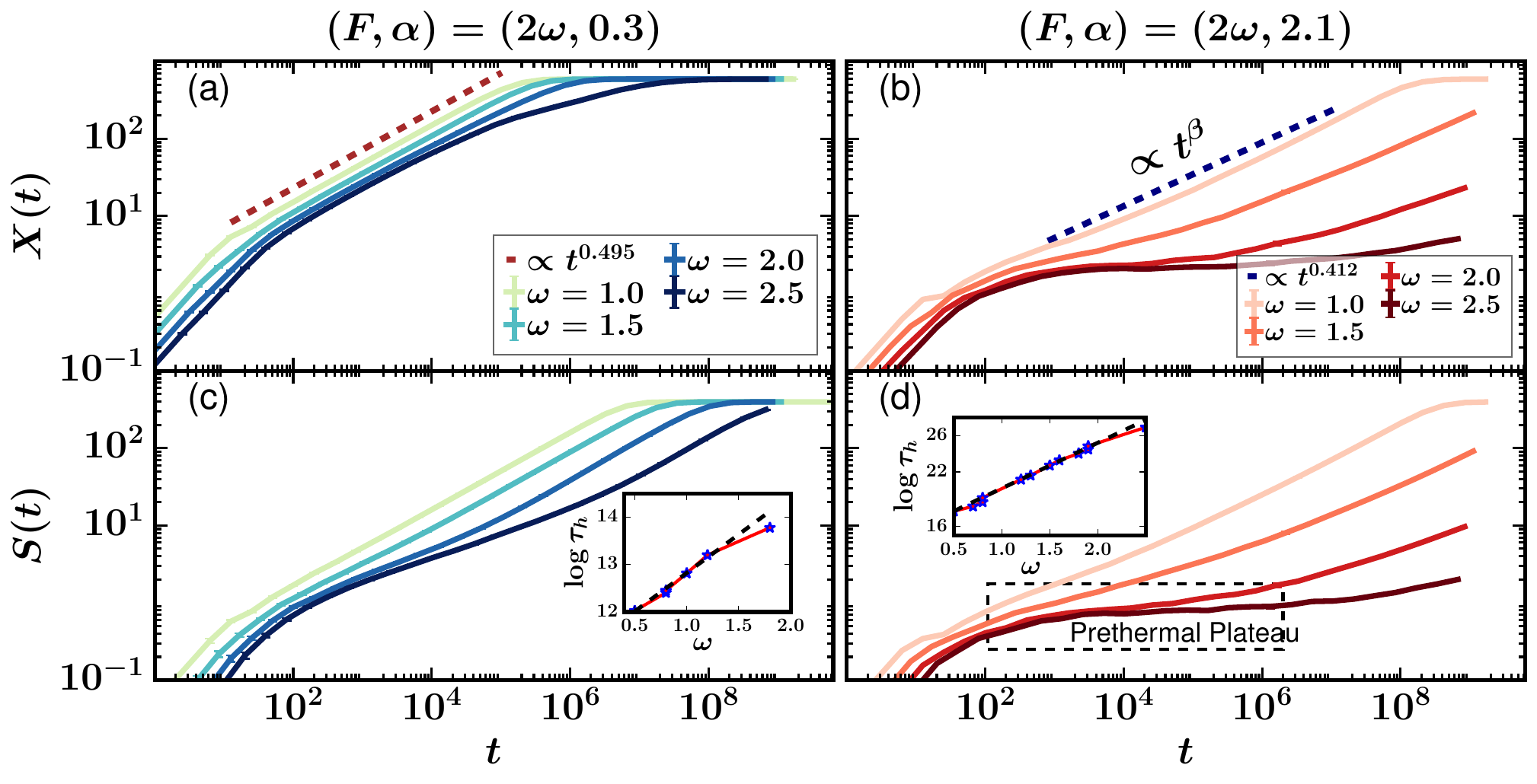}
	\includegraphics[width=0.34\textwidth]{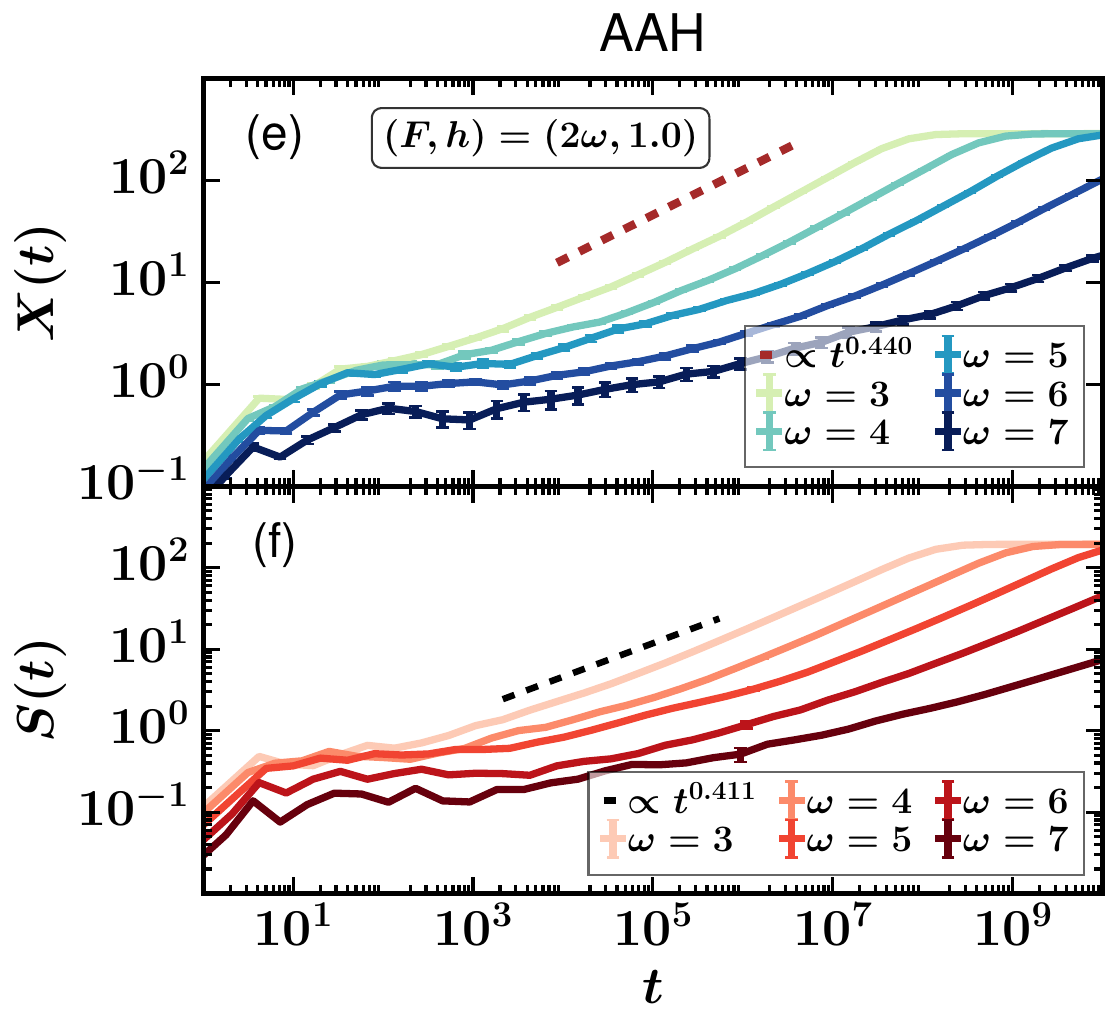}
	\caption{Dynamics of Thue-Morse driven system tuned at dynamical localization point ($F=2\omega$). (a,c) Dynamics of RMSD $X(t)$ for Thue-Morse driven PLRBM model for different frequencies $\omega$, and other parameters $\lbrace \alpha,  L\rbrace=\lbrace 0.3, 2048\rbrace $. The dashed line in (a) represents the power-law fit signifying diffusive transport, $t^{0.5}$. The inset in (b) shows heating time $\tau_{h}$ with driving-frequency $\omega$. (b,d). Dynamics of RMSD $X(t)$ and entanglement entropy $S(t)$ for Thue-Morse driven disordered system for different frequencies, and other parameters $\lbrace \alpha,  L\rbrace=\lbrace 2.1, 2048\rbrace $. The blue dashed line in (b) corresponds to subdiffusive transport ($t^{\beta}$). Inset in (d) exhibits heating-time dependence on driving frequency. (e,f). Dynamics of RMSD $X(t)$ and entanglement entropy $S(t)$ for Thue-Morse driven AAH model for disordered strength $h=1.0$ and $\lbrace J, L\rbrace=\lbrace 1, 2048\rbrace $. The dashed lines in (e,f) correspond to subdiffusive transport, sublinear growth of entanglement entropy, respectively. }
	\label{fig:fig4}
\end{figure*}
We further present the decay of $X_{sat}$ with long-range exponent
$\alpha$ for the drive-parameters tuned at the DL point
$F=2\omega$. The decay of $X_{sat}$ shows that as $\alpha$ increases,
the system approaches the short-range hopping limit where the
dynamical localization has already been
reported~\cite{tiwari2024dynamical}. Furthermore, the decrease in
$X_{sat}$ with system-size proves that the phenomenon of dynamical
localization observed in an aperiodically driven long-range system is
not a finite-size effect but valid in the thermodynamic limit
(Fig.~\eqref{fig:fig5b}(c)). The constant saturation value of
$X_{sat}$ at ADL point $F=1.6\omega$ confirms the complete
delocalization of the system irrespective of long-range exponent
(Fig.~\eqref{fig:fig5b}(b)). The monotonic increase of $X_{sat}$ with
$L$ again confirms the absence of localization at ADL points as shown
in Fig.~\eqref{fig:fig5a}(b).

Hence, this extends the notion of \textit{exact dynamical
  localization} (EDL) for piecewise discrete quasiperiodic driving
which was earlier known only for periodic discontinuous
driving~\cite{Dignam2002conditions}.

\subsection{Disordered long-range hopping} 
We now study the impact of the aperiodic Thue-Morse drive on the
disordered long-range hopping model ($J=0, u_{ij}\neq 0$)
(Eq.~\eqref{eqn:eq1a}). We focus on the delocalized ($\alpha=0.3$) and
localized ($\alpha=2.1$) phases of the undriven model and tune the
parameters at the dynamical localization point of the clean model
$F=2\omega$.

The dynamics of the root-mean squared displacement $X(t)$ and the
entanglement entropy $S(t)$ is plotted in Fig.~\ref{fig:fig4}(a-d) for
a range of driving frequencies $\omega=1-2.5$. In the delocalized
phase ($\alpha=0.3$), we observe diffusive transport where
$X(t)\propto t^{1/2}$ and eventually reaches to the
infinite-temperature value. In the localized phase ($\alpha=2.1$), on
the other hand, we observe different dynamical regimes: an initial
growth of $X(t)$ and $S(t)$, followed by a plateau whose width
increases with increasing the driving frequency, and an eventual
subdiffusive growth ($ X(t)\propto t^{\beta}$, $\beta<0.5 $) towards
the infinite-temperature value in the long-time limit. This behavior
can be intuitively understood by looking at the Fourier spectrum of
the Thue-Morse sequence which contains multiple frequencies (both low
and high). These multiple frequencies create several channels that
facilitate transitions between the energy levels of the undriven
model, and thus influence the dynamics of the system under periodic
driving. In the high-frequency regime, these channels are suppressed
upto a long time and thereby gives rise to the plateau behavior. In
the low-frequency regime, all the channels are active and hence lead
to immediate diffusion. To depict the dependence of heating time on
driving frequency, we compute the heating time as the time when the
entanglement entropy reaches half of the value corresponding to the
infinite temperature state. We show the data for $\tau_{h}$ vs
$\omega$ in the linear ($\omega-$axis)-log ($\tau_{h}$-axis) scale in
the inset of Fig.~\eqref{fig:fig4}(c,d) for $\alpha=0.3, 2.1$ which
suggests that the heating time follows exponential growth with driving
frequency, $\tau_{h}\propto \exp(\omega)$.

\begin{figure*}[ht!]
	\hspace{-3ex}\includegraphics[width=0.62\textwidth]{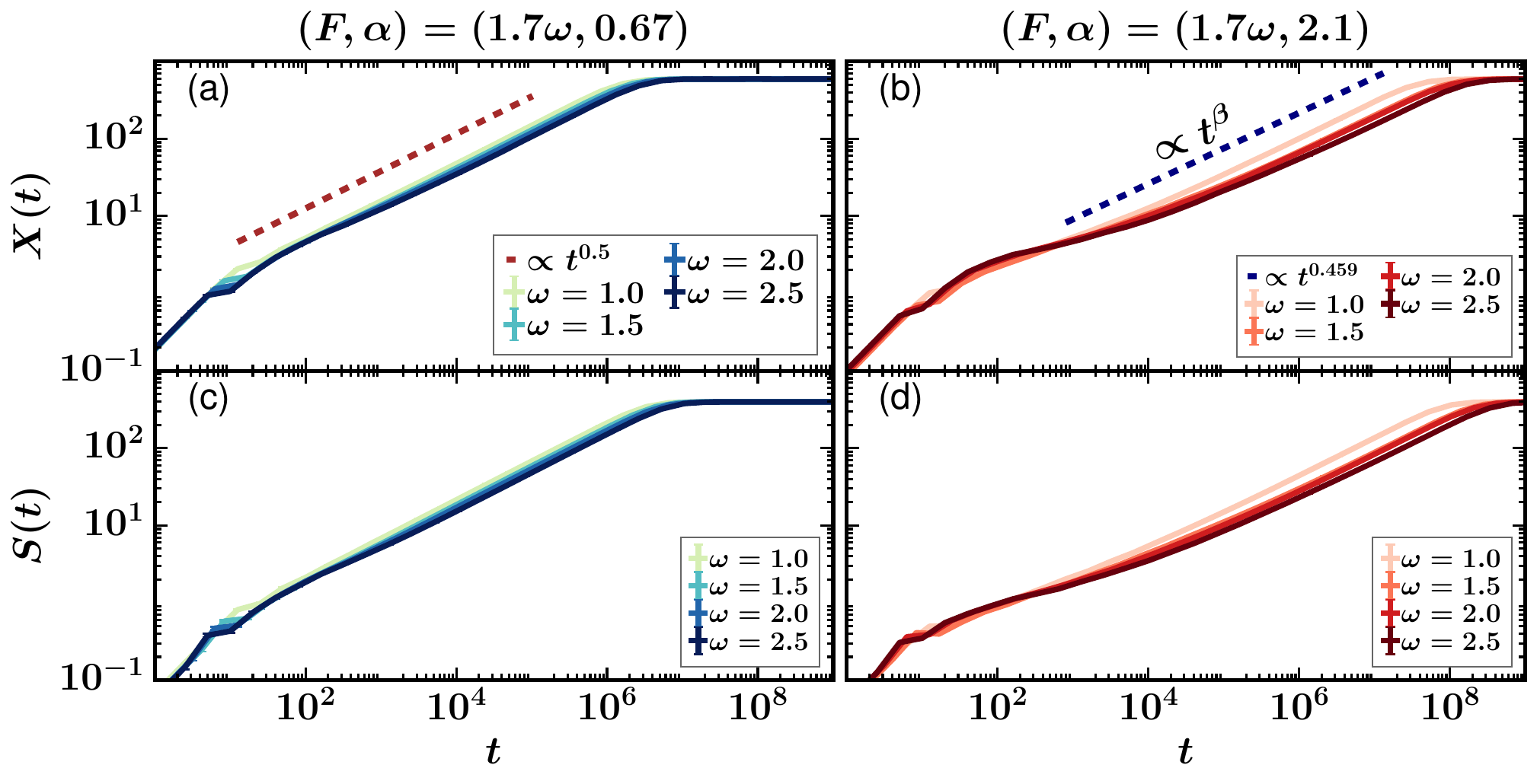}
	\includegraphics[width=0.34\textwidth]{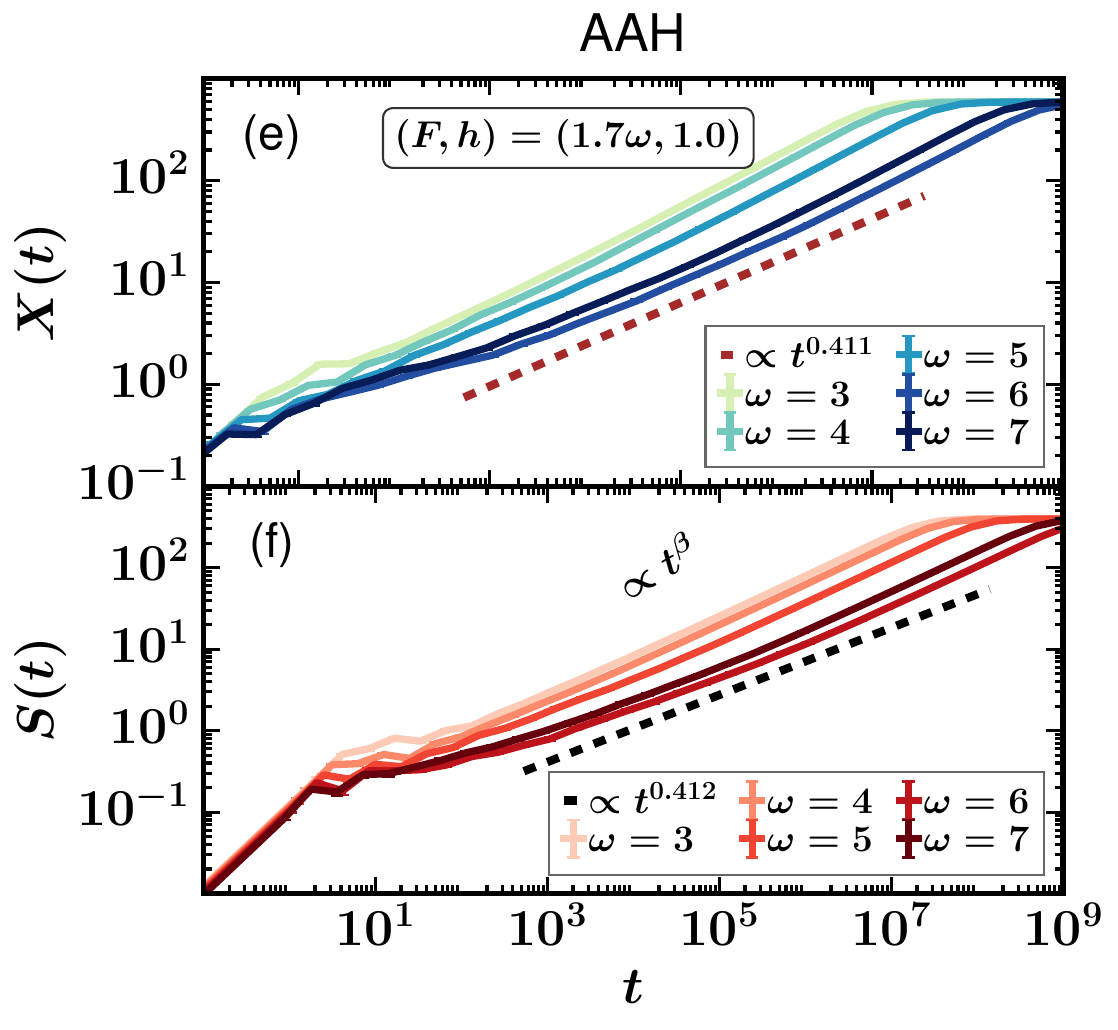}
	\caption{Dynamics of Thue-Morse driven system tuned at away from dynamical localization point ($F=1.7\omega $) for different frequencies. (a,b) Dynamics of RMSD $X(t)$ for Thue-Morse driven PLRBM model for different frequencies for long-range exponents $\alpha=0.67, 2.1$, system-size $L=2048$. The dashed lines in correspond to power-law fit $\propto t^{\beta}$. (c,d) Dynamics of entanglement entropy $S(t)$ for Thue-Morse driven disordered system for different frequencies $\omega $, and $\alpha=0.67, 2.1 $, system-size $L=2048$. (e,f) Dynamics of RMSD $X(t)$ and entanglement entropy $S(t)$ for Thue-Morse driven AAH model for disordered strength $h=1.0$ and $\lbrace J, L\rbrace=\lbrace 1, 2048\rbrace $. The dashed lines in (e,f) presents the power-law fit $\propto t^{\beta}$.}
	\label{fig:fig4b}
\end{figure*}

Similar observations can be seen in the dynamics of the entanglement
entropy which either grows directly to the Page value without
featuring a prethermal plateau in the delocalized phase and with a
prethermal plateau in the localized phase. For the tuning at ADL
points of the clean limit, Fig.~\eqref{fig:fig4b} shows that the
dynamics of transport features transition from diffusive to
subdiffusive on varying the long-range exponent, from $\alpha<1$ to
$\alpha>1$ similar to Fig.~\eqref{fig:fig4}. In comparison to the
tuning at DL points (Fig.~\eqref{fig:fig4}(b,d)), the dynamics of the
system tuned at ADL points does not feature any prethermal plateau as
shown in Fig.~\eqref{fig:fig4b}(b,d). The dependence on frequency at
DL tuning, and the independence on frequency for ADL tuning are also
visible from Figs.~\eqref{fig:fig4}(a-d) and ~\eqref{fig:fig4b}(a-d)),
respectively, similar to what we saw earlier for the periodically
driven PLRBM model.

\subsection{Comparison with quasiperiodic AAH model}
In this subsection, we contrast our results with a Thue-Morse driven short-range model which also features a delocalization to localization transition in its static limit. To this end, we focus on a quasi-periodic Aubry-Andr{\'e}-Harper model~\cite{aubry1980analyticity}. The Hamiltonain can be written as 
\begin{eqnarray}
\label{Ht_AAH}
 H(t)&=& H_{\text{AAH}}+\mathcal{F}(t)\sum_{j=0}^{L-1}jn_{j},\\
 \label{H_AAH}
	H_{\text{AAH}}&=& -J\sum_{j}\left(c_{j}^{\dagger}c_{j+1}+h.c.\right)+\sum_{j}h_{j}c_{j}^{\dagger}c_{j}.
\end{eqnarray}
Here $\mathcal{F}(t)$ is the time-aperiodic electric field, $J=1$ is
the hopping strength and $h_j = h\cos(2\pi\beta j + \phi)$ is the
on-site potential with $h$ being the strength of the potential, $\beta
= (\sqrt{5}-1)/2$ is an irrational number and $\phi$ is a global phase
that is being averaged over~\cite{
  aubry1980analyticity,harper1955single}. The AAH model
(Eq.~\eqref{H_AAH}) features a localization to delocalization
transition on varying the potential strength. For $h<2$, all the
eigenstates are delocalized; for $h>2$, all the eigenstates are
localized. The eigenstates at the transition point feature
multifractal behavior. The transport is ballistic in the delocalized
phase, anomalous at the transition point, and absent in the localized
phase~\cite{Varma2017fractality} thereby featuring a dynamical phase
transition from ballistic to no transport at $h=2.0J$.  As we show
ahead, the presence of the electric-field drive here affects the
system differently compared to the disordered long-range model
discussed in the previous sections.

In Fig.~\ref{fig:fig4}(e), we plot the dynamics of root mean squared
displacement $X(t)$ for a range of driving frequencies
$\omega=3.0-7.0$ and for $h=1.0$ corresponding to the delocalized
phase. As can be seen, the RMSD exhibits distinct transport regimes--
a plateau after an initial growth followed by subdiffusion to the
infinite temperature value ($\propto L$). The width of the plateau
depends largely on the driving frequency, and for large driving
frequencies, the subdifusion kicks in very late. This gives rise to
drive-induced slow dynamics where the drive leads to slow dynamics in
the delocalized phase. In contrast, in the absence of the aperiodic
drive, the transport is known to be ballistic. This feature is similar
to the previously reported drive-induced slow relaxation for the
interacting systems in the ergodic phase of the disordered
model~\cite{tiwari2024dynamical}; however, we show it in a simple,
non-interacting setup here.

We plot the entanglement entropy in Fig.~\ref{fig:fig4}(f). Similar to
the RMSD, the entanglement entropy also exhibits distinct dynamical
regimes: it grows in time initially, then saturates up to a plateau
value and then starts to grow as a power law $t^{\beta} (\beta<0.5)$
before reaching the Page value ~\cite{vidmar2017entanglement}
suggesting the drive-induced slow relaxation in the delocalized phase
in the high-frequency limit. The crossover time again is exponentially
large in the driving frequency. The same behavior is also expected for
the parameters considered in the localized phase. In
Fig.~\eqref{fig:fig4b}(e,f), we also present the dynamics of the
aperiodically driven AAH model where the drive-parameters are tuned at
ADL points($F=1.7\omega$), and where the disorder-strength is
$h=1$. Thus the transport properties seen here are similar to those of
the quasiperioidcally driven long-range hopping model.

Some insights about the above behavior can be gained by performing a
high-frequency expansion for the AAH model. To this end, we first
focus on just two cycles of the Thue-Morse sequence and calculate the
effective Hamiltonian as ~\cite{tiwari2024dynamical}
\begin{eqnarray}\label{eq10:Heff_dis}
H_{\text{eff}}&=& J_{\text{eff}}\lbrace \hat{K}e^{-iFT/4}+\hat{K}^{\dagger}e^{iFT/4}\rbrace+ D_{0}+H_\text{LRH}.\nonumber\\
\end{eqnarray}
Here, $D_{0}=\sum_{j}h_{j}c_{j}^{\dagger}c_{j}$ corresponds to the
quasiperiodic potential part, and $H_\text{LRH}$ correponds to the
long-range hopping terms (order of higher-powers of
$T$)~\cite{tiwari2024dynamical}, and can be ignored in the
high-frequency limit.  In this limit, the effective Hamiltonian
becomes $H_\text{eff}(J, h)\approx H(J, h/J_\text{eff})$, suggesting
that the effect of the drive is to suppress the hopping strength or
conversely increasing the effective on-site potential strength. For a
perfect periodic drive, which is just a repetition of a two-cycle
sequence, one can dynamically adjust the transition point, potentially
extending it towards the delocalized side and eventually leading to
high-frequency driving-induced
localization~\cite{martinez2006delocalization,bairey2017driving,tiwari2024dynamical,dotti2024measuring}.
For the aperiodic Thue-Morse sequence that contains both the low and
high-frequency components within its Fourier spectrum, the frequencies
exceeding the local bandwidth do not affect localization. However,
lower frequencies induce transitions between localized states,
resulting in a plateau in the dynamics before the onset of
low-frequency components, which ultimately lead to subdiffusion.

\section{Summary and Conclusion}
\label{Summary}

We investigate the impact of a periodic and aperiodic Thue-Morse drive
on a disordered long-range model, which exhibits a
delocalization-to-localization phase transition. For the periodic
drive, we analyze the properties of the Floquet operator and transport
dynamics to uncover drive-induced effects on the PLRBM model. Below we
summarise our key findings:

\begin{enumerate}[label=(\roman*).]

\item The analysis of the level spacing ratio of the quasienergy
  spectrum and the generalized inverse participation ratio obtained
  from the quasienergy eigenstates suggests weak delocalization on
  the delocalized side of the undriven PLRBM model ($\alpha<1$). The
  expression for the effective Hamiltonian explains the emergence of
  the weak multifractal and fractal phase as an effect of the renormalization of the
  effective hopping where suppression in the hopping plays an
  essential role leading to the phenomenon of \textit{drive-induced
    shortening of the range of the long-range model}.

\item The other side of the undriven PLRBM model remains unaffected,
  and shows localization as a result of the interplay between
  renormalization of hopping, and shorter-range of hopping
  ($\alpha>1$) evident from the analysis of the effective Hamiltonian.

\item The transition is indicated by static measures such as the level
  spacing ratio and the generalized inverse participation ratio. It is
  also accompanied by a change of transport properties from diffusive to subdifusive on the delocalized side of the PLRBM model, and
  to slow logarithmic on the localized side as shown by the dynamics of RMSD $X(t)$ and
  entanglement entropy $S(t)$.
\end{enumerate}

For the aperiodic Thue-Morse driven long-range clean system, we derive
an expression for the effective Hamiltonian. This also yields the
condition for \textit{exact dynamical localization} which is an effect
of the discontinuity of the drive-protocol that has a sharp jump in
the drive-amplitude~\cite{dunlap1986dynamic,Dignam2002conditions}.

\begin{enumerate}[label=(\roman*).]

\item At EDL points, the transport of the system ceases, while at away
  from dynamical localization (ADL) points, ballistic transport is
  observed, as corroborated by the dynamics of RMSD and entanglement
  entropy.

\item Driving a long-range disordered system on the delocalized side
  ($\alpha<1$) yields diffusive transport eventually reaching the
  infinite-temperature state. On the other hand, driving a short-range
  system ($\alpha>1$) exhibits a metastable prethermal plateau
  followed by subdiffusion to the infinite-temperature state.

\item We also present a comparative study of the aperiodically driven
  long-range hopping model with the aperiodically driven AAH
  model. Similar to the quasiperiodically driven long-range hopping
  model, the quasiperiodically driven AAH model also exhibits a
  prethermal plateau followed by \textit{drive-induced subdiffusive
    relaxation} to the infinite-temperature state. Interestingly, the
  prethermal plateau is also observed for the AAH model tuned on the
  delocalized side.

\end{enumerate}

Our work opens up new directions to explore the Floquet/quasi-Floquet
engineering of experimentally realizable systems with long-range
coupling. While here we focus on discrete time-dependent driving, it
would be interesting to explore the effect of continuous
time-dependent electric-field driving on the
delocalization-to-localization transition where the existence of EDL
is not
possible~\cite{dunlap1986dynamic,Dignam2002conditions,eckardt2009exploring}. Moreover,
it would also be worth exploring the interplay of interaction,
disorder, and driving and investigate the fate of Stark many-body
localization~\cite{stark2024duffin} in the interacting counterpart of
the driven PLRBM model.

\newpage
\begin{widetext}
\appendix

\section{}
\section*{Dynamical Localization in Square-wave Driven Long-range System}\label{A:Effective_Hamiltonian}

We consider a periodically driven system with power-law decay hopping given by the following Hamiltonian,

\begin{align}
\label{eqn:A1}
H(t)=-\sum_{i,j=1}^{L-2}\frac{J_{ij}}{|i-j|^{\alpha}}\left(c_{i}^{\dagger}c_{j}+ h.c.\right)  \pm F\text{Sgn}\left(\sin(\omega t)\right)\displaystyle \sum_{j=0}^{L-1}jn_{j}.
\end{align}
Here, $J_{ij}= J $ is the hopping strength,
and $\alpha$ is the long-range parameter. With the unitary transformation~\cite{hartmann2004dynamics} given by,
\begin{eqnarray}
  \label{eq:k}
  \hat{K}_{p}=\sum_{n}c_{n}^{\dagger} c_{n+p},\quad   \hat{N}=\sum_{n}nc_{n}^{\dagger}c_{n}.
\end{eqnarray}

\begin{eqnarray}
H(t)=-\sum_{p>0}\frac{J}{p^{\alpha}}\left(\hat{K}_{p}+\hat{K}_{p}^{\dagger}\right)\pm F Sgn(\sin(\omega t))\sum_{p=0}^{n-1}jn_{p},
\end{eqnarray}

where, $\hat{K}_{p}=\sum_{j}c_{j^{\dagger}}c_{j+p}, \quad \hat{K}_{p}^{\dagger}=\sum_{j}c_{j+p}^{\dagger}c_{j}, \quad \hat{N}=\sum_{j}j\hat{n}_{j}$, and $p=1,2,3,...(L-1)$. For the time-evoluation of the system, we construct time-evolution operators $U_{A}$ and $U_{B}$ corresponding to $H_{A}$ and $H_{B}$,

\begin{eqnarray}
U_{A}&=& \exp(-i(T/2)H_{A}), \quad U_{B}=\exp(-i(T/2)H_{B}),\\
H_{A, B}&=& -\sum_{p>0}\frac{J}{p^{\alpha}}\left(\hat{K}_{p}+\hat{K}_{p}^{\dagger}\right)\pm F \sum_{p=0}^{n-1}jn_{p}.
\end{eqnarray}

For $p=1$, it is well known nearest-neighbor hopping model where effective Hamiltonian can be written as

\begin{eqnarray}
H_{\text{eff}}^{[p=1]}&=& -\mathcal{J}_{\text{eff}}\left(\hat{K}e^{-iFT/4}+\hat{K}^{\dagger}e^{iFT/4}\right), \quad \mathcal{J}_{\text{eff}}=J\frac{\sin(FT/4)}{\left(FT/4\right)}
\end{eqnarray}

Similarly, we can evaluate the effective Hamiltonian for all $p'$s using BCH formalism,

\begin{eqnarray}
& H_{\text{eff}}^{[p=2]}= -\mathcal{J}_{\text{eff}}^{[p=2]}\left(\hat{K}_{2}e^{-2iFT/4}+\hat{K}^{\dagger}_{2}e^{2iFT/4}\right)&,\nonumber\\
&H_{\text{eff}}^{[p=3]}=-\mathcal{J}_{\text{eff}}^{[p=3]}\left(\hat{K}_{3}e^{-3iFT/4}+\hat{K}^{\dagger}_{3}e^{3iFT/4}\right)&,\nonumber\\
&..............................................&,\nonumber\\
& H_{\text{eff}}^{[p]}=-\mathcal{J}_{\text{eff}}^{[p]}\left(\hat{K}_{p}e^{-piFT/4}+\hat{K}^{\dagger}_{p}e^{piFT/4}\right),&
\end{eqnarray}

where $\mathcal{J}_{\text{eff}}^{[p]}= \frac{J\sin (pAT/4)}{p^{\alpha}(pAT/4)} $. Hence, we get an effective Hamiltonian,

\begin{eqnarray}
H_{\text{eff}}=-\sum_{p}\mathcal{J}_{\text{eff}}^{[p]}\left(\hat{K}_{p}e^{-piFT/4}+\hat{K}^{\dagger}_{p}e^{piFT/4}\right).
\end{eqnarray}

\section{}
\section*{Effective Hamiltonian For Periodically Driven Disordered System}\label{A:1Effective_Hamiltonian}

We consider a periodically driven system with nearest-neigbor disordered hopping amplitude given by following Hamiltonian,
\begin{eqnarray}
\label{Hf}
H(t)=-\sum_{n} J_{n}\left(c_{n}^{\dagger}c_{n+1}+H.C.\right)\pm F \text{Sgn}(\sin(\omega t))\sum_{n}^{L-1}nc_{n}^{\dagger}c_{n},
\end{eqnarray}

we can rewrite the Hamiltonian (Eq.~\eqref{Hf}) as

\begin{eqnarray}
\label{Hfa}
H(t)=-\sum_{n} J_{n}\left(\hat{K}_{1}+\hat{K}_{1}^{\dagger}\right)\pm F \text{Sgn}(\sin(\omega t))\sum_{n}^{L-1}\hat{N},
\end{eqnarray}

where, $J_{n}\in [-W, W]$, and from Eq.~\eqref{eq:k}, $\hat{K}_{1}=\sum_{n}c_{n^{\dagger}}c_{n+1}, \quad \hat{K}_{1}^{\dagger}=\sum_{n}c_{n+1}^{\dagger}c_{n}, \quad \hat{N}=\sum_{n}n\hat{n}_{n}$. 

For the time-evoluation of the system, we construct time-evolution operators $U_{A}$ and $U_{B}$ corresponding to $H_{A}$ and $H_{B}$,

\begin{eqnarray}
U_{A}&=& \exp(-i(T/2)H_{A}), \quad U_{B}=\exp(-i(T/2)H_{B}),\\
H_{A, B}&=& -\sum_{n} J_{n}\left(\hat{K}_{1}+\hat{K}_{1}^{\dagger}\right)\pm F \sum_{n}^{L-1}\hat{N}.
\end{eqnarray}

Using BCH expansion,

\begin{eqnarray}
U_{A}U_{B}&\equiv & U_{op}  \equiv \exp(-iTH_{\text{eff}}),  \\
\label{Hf_nn}
H_{\text{eff}}&=& -\sum_{n}J_{n}\left(\frac{\sin (FT/4)}{FT/4}\right)\left(c_{n}^{\dagger}c_{n+1}e^{-iFT/4}+h.c.\right)+ H_{1}
\end{eqnarray}
where
\begin{eqnarray}
\label{Eq.Hp1}
H_{1}&=& -\left(\frac{-iT}{2}\right)^{4}\left[\sum_{n}J_{n}^{2}\left(2J_{n}-J_{n+1}-J_{n-1}\right)\left(c_{n+1}^{\dagger}c_{n}-c_{n}^{\dagger}c_{n+1}\right)\right]+...... .
\end{eqnarray}

Eq.~\eqref{Hf_nn} shows that the driving the system with high-frequency preserves the dynamical localization in the system where $H_{1}$ can be ignored. However, driving with smaller frequency leads to the delocalization in the system where correction terms $H_{1}$ plays effective role.

Similarly, we can extend our understanding from  nearest-neighbor model to long-range hopping model which in the clean limit is known to show the phenomenon of exact dynamical localization. Following the formalism shown above, we can write an approximate effective Hamiltonian for a periodically driven PLRBM model defined as

\begin{eqnarray}
\label{Hf2}
H(t)=-\sum_{n} \left(\frac{J_{ij}}{|i-j|^{\alpha}}\hat{c}_{i}^{\dagger}\hat{c}_{j}+H.C.\right)\pm F Sgn(\sin(\omega t))\sum_{n}^{L-1}\hat{N},
\end{eqnarray}

Using BCH formalism, we write following effective Hamiltonian,

\begin{eqnarray}
\label{HF3}
H_{\text{eff}}&=& \sum_{ij} \left(\frac{J_{ij}\sin
  ((p)FT/4)}{p^{\alpha}(pFT/4)}e^{-piFT/4}\hat{c}_{i}^{\dagger}\hat{c}_{j}+H.C.\right)+H_{1},\quad (p=|i-j|).
  \end{eqnarray}
We define the unitary operators as follows,
\begin{eqnarray}
  \label{eq:k}
  \hat{K}_{p}=\sum_{n}c_{n}^{\dagger} c_{n+p},\quad   \hat{N}=\sum_{n}nc_{n}^{\dagger}c_{n},\quad \left(\mathcal{J}_{\text{eff}^{[p]}}=\frac{J_{ij}\sin (pFT/4)}{p^{\alpha}(pFT/4)}\right),
\end{eqnarray}  
and effective Hamiltonian can again be expressed as
\begin{eqnarray}  
\label{Hf_ed}
H_{\text{eff}}&=&-\sum_{p}\mathcal{J}^{[p]}_{\text{eff}}\left(\hat{K}_{p}e^{-piFT/4}+\hat{K}^{\dagger}_{p}e^{piFT/4}\right)+H_{1},
\end{eqnarray}

\begin{eqnarray}
H_{1}=-\left(\frac{-iT}{2}\right)^{4}\left[\sum_{n}J_{n}^{2}\left(2J_{n}-J_{n+1}-J_{n-1}\right)\left(c_{n+1}^{\dagger}c_{n}-c_{n}^{\dagger}c_{n+1}\right)+\sum_{n}V_{n}\left(c_{n}^{\dagger}c_{n+3}+H.C.\right)\right]+...... .
\end{eqnarray}

Eq.~\eqref{Hf_ed} shows that at high frequencies, the system is predominantly governed by the first term of the effective Hamiltonian, corresponding to the PLRBM model with a renormalized hopping strength. Higher-order terms in the power of the time period 
$T$ can be neglected in this regime.
\end{widetext}

\bibliography{refv_new}

\end{document}